\documentclass[10pt,journal,twocolumn]{IEEEtran}
\IEEEoverridecommandlockouts
\usepackage{cite}
\usepackage{amsmath,amssymb,amsfonts,booktabs}
\usepackage{algorithmic}
\usepackage{graphicx}
\usepackage{textcomp}
\usepackage{xcolor}
\usepackage{enumerate}
\usepackage{epstopdf}
\usepackage{amsthm}

\usepackage{algorithm}
\usepackage{algorithmic}
\allowdisplaybreaks[4]

\usepackage{array}
\usepackage{stfloats}
\usepackage{cuted}
\usepackage{url}
\usepackage{verbatim}

\newtheorem{theorem}{Theorem}
\newtheorem{remark}{Remark}
\newtheorem{proposition}{Proposition}
\newtheorem{assumption}{Assumption}
\usepackage{color}
\begin{document}
	







\title{Latency Minimization for UAV-Enabled\\Federated Learning: Trajectory Design\\and Resource Allocation}

\author{

        Xuhui Zhang,
        Wenchao Liu,
        Jinke Ren,
        Huijun Xing,
        Gui Gui,
        Yanyan Shen,
        and Shuguang Cui

\thanks{
X. Zhang and J. Ren are with the Shenzhen Future Network of Intelligence Institute (FNii-Shenzhen), the School of Science and Engineering (SSE), and the Guangdong Provincial Key Laboratory of Future Networks of Intelligence, The Chinese University of Hong Kong, Shenzhen, Guangdong 518172, China (e-mail: xu.hui.zhang@foxmail.com; jinkeren@cuhk.edu.cn).
(\emph{Corresponding author: Jinke Ren})
}

\thanks{
W. Liu is with Shenzhen Institute of Advanced Technology (SIAT), Chinese Academy of Sciences (CAS), Guangdong 518055, China, and also with Southern University of Science and Technology, Guangdong 518055, China (e-mail: wc.liu1@siat.ac.cn).
}

\thanks{
H. Xing is with the Department of Electrical and Electronic Engineering, Imperial College London, London SW7 2AZ, The United Kingdom (e-mail: huijunxing@link.cuhk.edu.cn).
}

\thanks{
G. Gui is with the School of Automation, Central South University, Hunan 410083, China (email: guigui@csu.edu.cn).
}

\thanks{
Y. Shen is with the SIAT, CAS, Guangdong 518055, China, and also with Shenzhen University of Advanced Technology, Guangdong 518055, China (e-mail: yy.shen@siat.ac.cn).
}

\thanks{
S. Cui is with the SSE, the FNii-Shenzhen, and the Guangdong Provincial Key Laboratory of Future Networks of Intelligence, The Chinese University of Hong Kong, Shenzhen, Guangdong 518172, China (e-mail: shuguangcui@cuhk.edu.cn).
}

}

\maketitle

\begin{abstract}

Federated learning (FL) has become a transformative paradigm for distributed machine learning across wireless networks.
However, the performance of FL is often hindered by the unreliable communication links between resource-constrained Internet of Things (IoT) devices and the central server. To overcome this challenge, we propose a novel framework that employs an unmanned aerial vehicle (UAV) as a mobile server to enhance the FL training process. By capitalizing on the UAV's mobility, we establish strong line-of-sight connections with IoT devices, thereby enhancing communication reliability and capacity. To maximize training efficiency, we formulate a latency minimization problem that jointly optimizes bandwidth allocation, computing frequencies, transmit power for both the UAV and IoT devices, and the UAV's flight trajectory.
{Subsequently, we analyze the required rounds of the IoT devices training and the UAV aggregation for FL convergence. Based on the convergence constraint, we transform the problem into three subproblems and develop an efficient alternating optimization algorithm to solve this problem effectively.}
Additionally, we provide a thorough analysis of the algorithm's convergence and computational complexity. Extensive numerical results demonstrate that our proposed scheme not only surpasses existing benchmark schemes in reducing latency {up to 15.29\%}, but also achieves training efficiency that nearly matches the ideal scenario.

\end{abstract}
\begin{IEEEkeywords}
Unmanned aerial vehicles, federated learning, resource allocation, trajectory design.
\end{IEEEkeywords}

\section{Introduction}
\IEEEPARstart{U}{nmanned} aerial vehicles (UAVs), also known as drones, are pivotal to sixth-generation (6G) networks and the low-altitude economy \cite{9468714, 10681882, 10693833},
offering maneuverability, flexible deployment, and operational efficiency for delivering communication-computation services in remote or infrastructure-limited areas \cite{9456851}.
Their applications span military operations, disaster response, medical logistics, and environmental monitoring \cite{8660516}.

{
While UAVs have demonstrated significant potential in diverse applications, their seamless integration into 6G wireless networks necessitates advanced strategies to address emerging challenges, such as dynamic resource allocation, stringent latency requirements, and data privacy concerns. To address these issues, machine learning (ML) has emerged as an effective technology \cite{9779322}. Specifically,}
the deployment of ML algorithms in wireless networks has attracted widespread attention in recent years \cite{8714026}.
Internet of Things (IoT) devices can solve classical wireless problems utilizing advanced ML algorithms, such as communication resource allocation, task offloading, and antenna beamforming design \cite{7792374, 9206115}.  
However, traditional ML algorithms request all devices to upload their local datasets to a remote cloud server for centralized training, which results in high communication overhead and severe privacy breach.
multiple distributed
Unlike traditional centralized ML requiring raw data uploads, federated learning (FL) empowers distributed IoT devices to collaboratively train shared models via parameter exchanges with a central server, inherently reducing communication overhead and preserving data privacy \cite{8666641}.
{These synergistic capabilities motivate our exploration of UAV-enabled FL systems, leveraging aerial agility for adaptive coverage and privacy-preserving ability of FL to tackle the communication-computation dilemmas in 6G networks.}

Despite the great potential of FL, its implementation in wireless networks still face a key challenge of unstable transmission quality \cite{9060868}. Specifically,
traditional FL systems take a 
base station (BS) as the central server to aggregate the model parameters from IoT devices. However, the communication links between IoT devices and the BS may experience blockages, making it challenging for IoT devices to complete local model transmission within a tolerant time threshold.
Fortunately, due to the flexible maneuverability and the capability to establish short-distance line-of-sight (LoS) communication links, the UAVs can serve as the central server to enable model aggregation and downloading in FL.
Nevertheless, both the UAVs and the IoT devices are typically energy-constrained, while the whole flight and training process require a continuous consumption of energy.
Therefore, it is important to optimize the UAV flight trajectory and communication resource allocation to meet the demands of FL training.

In this paper, we consider a UAV-enabled FL system, where the ground IoT devices work together with the UAV to train the FL model through wireless links under orthogonal frequency division multiple access (OFDMA).
We jointly optimize the trajectory of the UAV and the allocations of bandwidth, computation resources, and transmit power of both the UAV and IoT devices.
The main contribution of this article can be summarized as follows:
\begin{itemize}
    \item We propose a novel UAV-enabled FL system with OFDMA, where a UAV is dispatched as the central server to train the FL algorithm with multiple ground IoT devices during the flight time.
    Moreover, we formulate a latency minimization problem to improve the training efficiency of FL.
    \item To solve the latency minimization problem, we first analyze the convergence of the FL algorithm and then propose an alternating optimization (AO)-based algorithm.
    Furthermore, we analyze the convergence and computational complexity of the proposed scheme.
    \item Numerical results demonstrate the convergence, the total latency, and the training accuracy of the proposed AO-based algorithm.
    It is validated that
    the proposed scheme outperforms other benchmark schemes in terms of the latency and training accuracy, and
    always achieves the nearest performance to the ideal FL scheme.
\end{itemize}

The remainder of this paper is organized as follows.
Section II reviews the related works.
In Section III, we introduce the UAV-enabled FL system and formulate the latency minimization problem.
In Section IV, we analyze the FL convergence and propose the AO-based algorithm with convergence and computational complexity analysis.
Section V evaluates the performance of the proposed algorithm.
Finally, Section VI concludes the paper.

\section{Related Works}
\subsection{UAV-enabled Wireless Networks} 
UAV-enabled wireless systems
have high efficiency, reliable robustness, and flexible  motility \cite{8663615, 8489918, 8807386}.
Specifically,
Zeng \textit{et al.} \cite{8663615} studied a UAV energy consumption minimization problem to improve the communication throughput of mobile users.
Xie \textit{et al.} \cite{8489918} investigated a minimum throughput maximization problem for all ground users in a wireless powered communication network by optimizing the mobility of the UAV and the wireless resource allocation.
Cui \textit{et al.} \cite{8807386} studied the dynamic resource allocation with multiple UAVs by utilizing a deep reinforcement learning algorithm.
Moreover, the integration of the UAVs with other key technologies, such as mobile edge computing (MEC) \cite{8445936, 8956055}, space-air-ground integrated networks \cite{Zhang2023Learning, 10458883}, movable-antennas \cite{10654366}, intelligent reflecting surfaces / reconfigurable intelligent surface (IRS/RIS) \cite{9893192, 10098733}, IoT \cite{9273074, 9712630}, non-orthogonal multiple access systems \cite{9130430, 10214196}, blockchain \cite{10485478}, integrated sensing and comminations \cite{10233771, 10566041},
has significantly enhance the performance of communication systems. These motivate us to introduce the UAVs into FL systems to improve both the communication efficiency and training accuracy of FL.

\subsection{Wireless FL}
Recently, many works have studied the resource allocation, learning convergence, and communication efficiency issues in wireless FL systems \cite{9793704, 9210812, 9252924, 9264742, 9170917, 9606731, 10032291}.
Specifically,
Guo \textit{et al.} \cite{9793704} proposed a joint device selection and power control scheme for the wireless FL system, where the uplink and downlink communication resources were jointly optimized to minimize the performance gap between the expected and optimal global loss values.
Chen \textit{et al.} \cite{9210812} investigated the training of FL algorithms over wireless networks, where the communication resource allocation and user selection were jointly optimized to ensure the FL convergence.
Ren \textit{et al.} \cite{9252924} proposed a joint batchsize selection and communication resource allocation strategy for enhancing learning efficiency in a wireless FL system.
Yang \textit{et al.} \cite{9264742} studied the energy efficient model transmission and computing resource allocation to minimize the total energy consumption of the wireless FL system.
Ren \textit{et al.} \cite{9170917} proposed a novel scheduling policy in an FL system to exploit the optimal trade-off between channel quality and model importance.
Cao \textit{et al.} \cite{9606731} utilized over-the-air computation to enhance the communication efficiency in an FL system, where a novel power control was designed to maximize the convergence speed.
Wang \textit{et al.} \cite{10032291} proposed the IRS/RIS assisted over-the-air FL to enhance the data rates, and developed a graph neural network based algorithm to determine the channel coefficients.
Although the FL algorithm can be trained over wireless networks, the communication is the bottleneck when the channel condition is poor.

\subsection{UAV-supported FL} To overcome the communication bottleneck in FL, several works have utilized the UAV as the central server or mobile relay, to provide LoS links and enhance communication quality \cite{9210077, 9292475, 9374105, 10220154, 10051719, 10500333}.
Particularly,
Zhang \textit{et al.} \cite{9210077} studied the image classification task in a UAV-aided FL system, where a high classification accuracy at relatively low communication cost was achieved.
Ng \textit{et al.} \cite{9292475} utilized the UAVs as mobile relays to facilitate the communications between the vehicles and the server.
Pham \textit{et al.} \cite{9374105} proposed a UAV-assisted FL system, where the communication resource allocation and the UAV placement were jointly optimized to improve the UAV energy efficiency.
Fu \textit{et al.} \cite{10220154} optimized the total training time of FL by jointly optimizing the device scheduling, the UAV trajectory, and time allocation.
UAV-enabled FL systems empowered by convert communications were studied in \cite{10051719, 10500333}, where the UAVs were capable of imposing interfere to the eavesdroppers to further enhance the data security.
{However, the joint optimization of UAV trajectory design, bandwidth allocation, computation resource allocation, and transmit power control for both UAV and IoT devices remains underexplored. Existing studies often address these issues separately, failing to achieve the communication-computation-energy trade-off inherent in dynamic air-to-ground networks. Emerging UAV-enabled FL systems demand efficient resource coordination frameworks that holistically optimize these key parameters to ensure energy-efficient FL training while adhering to real-world deployment constraints, such as limited UAV onboard energy, heterogeneous IoT device capabilities, and time-varying channel conditions.}

\section{System Model}
\label{systemmodel}
\begin{figure}[!htbp]
	\centering
	\includegraphics[width=1\linewidth]{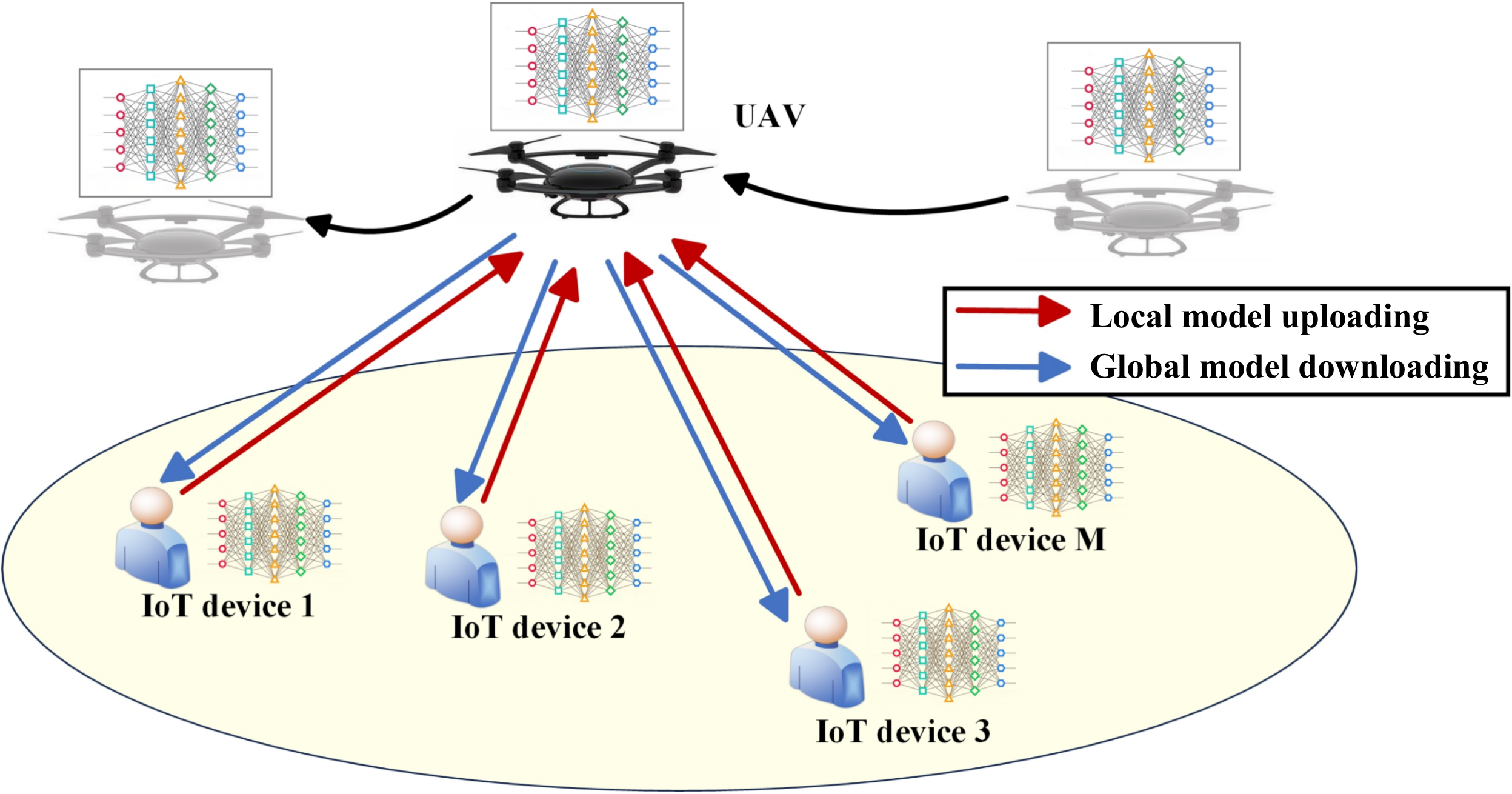}
	\caption{The UAV-enabled FL system.}
	\label{fig:system model}
\end{figure}

As illustrated in Fig. \ref{fig:system model}, we consider an FL system that consists of a UAV and $M$ IoT devices.
{
Each IoT device continuously collects data to update its local dataset, thereby serving certain privacy-preserving applications, such as surveillance video analysis, and face recognition.}
The UAV is equipped with a computing processor for model aggregation, serving as a mobile central server.
Each IoT device trains a local ML model and uploads its local model parameters to the UAV.
The set of IoT devices is denoted as $\mathcal{M} = \{ 1, 2, \cdots, M \}$. All devices are fixed within a geographical region of $D \times D\ {\rm{km}}^2$. Let $\mathcal{S}_m = (\mathbf{s}_{m}, 0)$ denote the location of the $m$-th IoT device, where $\mathbf{s}_m[n] = (x_m[n] , y_m[n] )$ denotes the two-dimensional (2D) Cartesian coordinates.

We consider a time block that is divided into $N$ time slots. In each time slot, the UAV needs to perform a round of training, including local model aggregation and global model downloading. The set of time slots is defined as $\mathcal{N} = \{ 1, 2, \cdots, N \}$.
Throughout the entire time block, the UAV continues to fly. Therefore, at the beginning of each time slot, we select a denotative position for the UAV to denote its current position. Let ${\mathcal{Q}}[n] = (\mathbf{q}[n], H)$ denote the position of the UAV at the $n$-th time slot, where $\mathbf{q}[n] = ( x[n] , y[n] )$ denotes the 2D Cartesian coordinates and $H_0$ is the fixed flying height.
{To enhance readability, the primary notations and their corresponding descriptions are given in Table \ref{tnoa}.}

\begin{table}[!htbp] \footnotesize
	\centering
	\caption{Notations and Descriptions}
	\label{tnoa}
	\begin{tabular}{c>{\centering\arraybackslash}p{6.825cm}}	
  \toprule	
  Notations& Description\\	
  \midrule  %
    $m,M,\mathcal{M}$ & The index, number, and set of the IoT devices\\ 
  $n,N,\mathcal{N}$ & The index, number, and set of time slots\\
    $u,U,\mathcal{U}$ & The index, number, and set of the RBs\\
  $\mathbf{q}[n]$ &  The 2D position of the UAV at time slot $n$\\
  $\mathbf{s}_m$ & The 2D location of the $m$-th IoT device\\
  $D_m[n]$ & The size of dataset at the $m$-th IoT device\\
  $\mathcal{D}_m[n]$ & The dataset at the $m$-th IoT device\\
  $\phi_m$ & The computational resources required for $1$ bit data\\
  $\vartheta$ & The data size of one data sample\\
  $R_m$ & The number of local training rounds of the $m$-th IoT device\\
  $d_m [n]$ & The size of the local model updates of the $m$-th IoT device\\
    $\rho_{\mathrm{sensed}}$ & The Poisson 
 density of sensed data\\
    $ \rho_{\mathrm{dropped}}$ & The Poisson density of dropped data\\
    $\boldsymbol{\theta}_m [n]$ & The local model at the $m$-th IoT device\\
    $\boldsymbol{\theta} [n]$ & The global model at the UAV\\
    $f_m [n]$ & The computing frequency of the $m$-th IoT device\\
    $f_{\max}$ & The maximum computing frequency of IoT devices\\
    $T_m^{\mathrm{train}} [n]$ & The local training latency of the $m$-th IoT device\\
    $E_m^{\mathrm{train}} [n]$ & The energy consumption for training of the $m$-th IoT device\\
    $g_m [n]$ & The channel gain from the $m$-th IoT device to the UAV\\
    $\alpha_{m, u} [n]$ & The allocation indicator of the $u$-th RB to the $m$-th IoT device\\
    $R_m^{\rm{up}} [n]$ & The data rate for
    model uploading of the $m$-th IoT device\\
    $T_m^{\rm{up}} [n]$ & The latency for
    model uploading of the $m$-th IoT device\\
    $E_m^{\rm{up}} [n]$ & The energy consumption for
    model uploading of the $m$-th IoT device\\
    $T_m^{\mathrm{wait}} [n]$ & The waiting latency of the $m$-th IoT device\\
    $f_{\mathrm{UAV}} [n]$ & The computing frequency of the UAV\\
    $F_{\max}$ & The maximum computing frequency of the UAV\\
    $T_{\mathrm{UAV}}^{\mathrm{agg}} [n] $ & The latency for model aggregation\\
    $E_{\mathrm{UAV}}^{\mathrm{agg}} [n] $ & The energy consumption for model aggregation\\
    $T_m^{\mathrm{down}} [n]$ & The latency of global model downloading for the $m$-th IoT device\\
    $R_m^{\mathrm{down}} [n]$ & The model downloading rate from the UAV to the $m$-th IoT device\\
    $E_m^{\mathrm{down}} [n] $ & The energy consumption for model downloading to the $m$-th IoT device\\
    $T_{\mathrm{FL}} [n]$ & The overall latency of the $n$-th training round\\
    $L_m[n]$ & The latency of the $m$-th IoT device\\
		\bottomrule
	\end{tabular}
\end{table}

\subsection{FL Model}
Each round of FL training consists of four steps, including local training at each IoT device, local model uploading from each IoT device to the UAV, model aggregation at the UAV, and global model downloading by the UAV.
Meanwhile, each local training round is completed within one time slot.
Let $\zeta_m [n] = \langle D_m[n], \phi_m, \vartheta, R_m, d_m[n] \rangle$ denote the local training task of the $m$-th IoT device, where $D_m[n]$ denotes the local dataset size in time slot $n$, $\phi_m$ denotes the computational resources required to process one bit of data, $\vartheta$ is the data size of one data sample, $R_m$ denotes the number of local training rounds, and $d_m [n]$ denotes the size of the local model updates in time slot $n$.
Specifically, the dataset at the $m$-th IoT device is given by
\begin{equation}
    \mathcal{D}_m[n]=\{(\boldsymbol{x}_{m}^1[n], y_m^1[n]),\cdots,(\boldsymbol{x}_{m}^{\vert \mathcal{D}_m[n] \vert}[n], y_m^{\vert \mathcal{D}_m[n] \vert}[n])\},
\end{equation}
where ${\vert \mathcal{D}_m[n] \vert} = \frac{D_m[n]}{\vartheta}$ denotes the number of data samples at the $m$-th IoT device, $\boldsymbol{x}_{m}^i[n] \in \mathbb{R}^L$ is the $i$-th data sample, and $y_{m}^i[n] \in \mathbb{R}$ represents the label related to $\boldsymbol{x}_{m}^i[n]$.

We assume that each IoT device has the capability to sense data from its environment to update its local dataset \cite{10214035}. The number of newly sensed data samples follows a Poisson distribution $\mathcal{P}$ with density $\rho_{\mathrm{sensed}}$, i.e., $D_m^{\mathrm{sensed}} [n] = \mathcal{P} (\rho_{\mathrm{sensed}} ) [n]$ data samples sensed in time slot $n$. Meanwhile, the old data of each IoT device may drop out, where the number of dropped old data samples follows a Poisson distribution with density $\rho_{\mathrm{dropped}}$, i.e., $D_m^{\mathrm{dropped}} [n] = \mathcal{P} (\rho_{\mathrm{dropped}} ) [n]$ data samples dropped in time slot $n$. Therefore, the number of training data samples in time slot $n$ is given by
\begin{equation}
    {\vert \mathcal{D}_m[n] \vert}  = {\vert \mathcal{D}_m[n-1] \vert}  + D_m^{\mathrm{sensed}} [n] - D_m^{\mathrm{dropped}} [n].
\end{equation}
To ensure local training at each IoT device, we assume that the training dataset satisfies
$D_{\mathrm{lb}} \le {\vert \mathcal{D}_m[n] \vert}  \le D_{\mathrm{ub}}$, where $D_{\mathrm{lb}}$ and $D_{\mathrm{ub}}$ 
denote the lower and upper bounds of the number of training samples.

\paragraph{Local training}
After receiving the global model at the last time slot, each IoT device trains its local model according to current local dataset. Let $\boldsymbol{\theta}_m [n]$ and $\mathcal{L}(\boldsymbol{\theta}_m [n]; \mathcal{D}_m [n])$ denote the local model and loss function of the $m$-th IoT device on its local dataset $\mathcal{D}_m [n]$, respectively. Then the loss function is given by
\begin{equation}
    \mathcal{L}(\boldsymbol{\theta}_m [n]; \mathcal{D}_m [n]) = 
    \frac{1}{\vert \mathcal{D}_m[n] \vert}\sum_{i=1}^{\vert \mathcal{D}_m[n] \vert} \mathcal{L}(\boldsymbol{\theta}_m [n]; \boldsymbol{x}_m^i [n], y_m^i [n]).
\end{equation}
Let $\boldsymbol{\theta} [n]$ denote the global global model, and $\Tilde{\boldsymbol{\theta}}_m [n]$ denote the update model by the $m$-th IoT device.
The local optimization problem to update $\Tilde{\boldsymbol{\theta}}_m [n]$ can be given by
\begin{equation}
\begin{split}
    &\min_{\Tilde{\boldsymbol{\theta}}_m [n]}
    \Tilde{\mathcal{L}}(\boldsymbol{\theta}_m [n]; \mathcal{D}_m [n]) \triangleq
    \mathcal{L}(\boldsymbol{\theta}_m [n]; \mathcal{D}_m [n]) \\&\quad -
    \left(
        \nabla \mathcal{L} (\boldsymbol{\theta} [n]; \mathcal{D}_m [n]) - \varrho \nabla \mathcal{L} (\boldsymbol{\theta} [n]; \mathcal{D}_m [n])
    \right)^{\mathsf{T}}
    \Tilde{\boldsymbol{\theta}}_m [n],
\end{split}
\label{localob}
\end{equation}
where $\varrho$ is a constant value. The gradient is utilized to update problem \eqref{localupdate}, which is given by
\begin{equation}
    \Tilde{\boldsymbol{\theta}}_m^r [n] = 
    \Tilde{\boldsymbol{\theta}}_m^{r-1} [n] -
    \upsilon \nabla \Tilde{\mathcal{L}}(\boldsymbol{\theta}_m [n]; \mathcal{D}_m [n]),
    \label{localupdate}
\end{equation}
where $r \in \{ 1,2,\cdots, R_m\}$ denotes the current local training round, and $\upsilon$ is the learning rate.
\paragraph{Local model uploading}
Each IoT device uploads the local model parameters to the UAV through wireless transmission.
The details of local model uploading will be presented at Section \ref{systemmodel}-B.
\paragraph{Model aggregation}
The training problem at the UAV can be formulated as
\begin{equation}
    \min_{\boldsymbol{\theta}[n]} \mathcal{L}(\boldsymbol{\theta}[n]) \triangleq
\sum_{m=1}^M \frac{\vert \mathcal{D}_m[n]\vert}{\sum_{m=1}^M 
\vert \mathcal{D}_m[n] \vert}
\mathcal{L}(\boldsymbol{\theta}_m [n]; \mathcal{D}_m [n]).
\end{equation}
When the UAV receives the local models from all IoT devices, it performs model aggregation by
\begin{equation}
    \boldsymbol{\theta} [n+1] = \boldsymbol{\theta} [n] + 
    \frac{1}{M}\sum_{m=1}^M \Tilde{\boldsymbol{\theta}}_m^{R_m} [n].
\label{uavaggre}
\end{equation}

\paragraph{Model downloading}
The UAV downloads the aggregated global model to all IoT devices through wireless transmission.
The details of global model downloading will be presented at Section \ref{systemmodel}-C.

\subsection{Local Update and Uploading Model}
Let $f_m [n]$ denote the computing frequency of the $m$-th IoT device in time slot $n$, {which is constrained by the maximum computing frequency of IoT devices $f_{\max}$, i.e., $f_m [n] \leq f_{\max}$.}
Then, the local training latency of the $m$-th IoT device in time slot $n$ is expressed as
\begin{equation}
T_m^{\mathrm{train}} [n] = \frac{D_m[n]\phi_m}{f_m [n]} R_m.
\end{equation}
The energy consumption for local model training can be expressed as
\begin{equation}
E_m^{\mathrm{train}} [n] = \kappa_{m} (f_m [n])^3 T_m^{\mathrm{train}} [n],
\end{equation}
where $\kappa_{m}$ denotes the effective capacitance of the $m$-th IoT device \cite{8488502, 9149631}.

Upon completing local training, each IoT device uploads the local model parameters to the UAV. The channels between the UAV and IoT devices are modeled as line-of-sight (LoS) channels since the UAV is flying at a relatively high altitude.
Let $g_m [n]$ denote the channel power gain from the $m$-th IoT device to the UAV in time slot $n$, which can be expressed as 
\begin{equation}
g_m [n] = \frac{\rho_{0}}{(\Delta_m[n])^{2}} = \frac{\rho_{0}}{H_{0}^{2} + \|\mathbf{q} [n] - \mathbf{s}_m\|^{2}},
\end{equation}
where $\rho_{0}$ denotes the channel power at the reference distance $d_{0} = 1 \text{m}$, and $\Delta_m[n]$ denotes the distance from the $m$-th IoT device to the UAV in time slot $n$.
Moreover, we adopt the OFDMA technology to allocate a total of $U$ orthogonal resource blocks (RBs) to IoT devices. These $U$ RBs are represented by the set $\mathcal{U} = \{ 1, 2, \cdots, U \}$. 
Let $\alpha_{m, u} [n] = \{0,1\}$ denote the allocation indicator of the $u$-th RB to the $m$-th IoT device in time slot $n$. Specifically, $\alpha_{m, u} [n] = 1$ denotes that the $u$-th RB is allocated to the $m$-th IoT device, otherwise $\alpha_{m, u}[n] = 0$.
Therefore, the data rate for the local model uploading can be expressed as
\begin{equation}
R_m^{\rm{up}} [n] = \sum_{u=1}^{U} \alpha_{m, u} [n] B \log_2 \left(1 + \frac{p_m[n] g_m[n]}{ B \sigma^2}\right),\label{rate_up}
\end{equation}
where $B$ denotes the bandwidth of a single RB, $p_m [n]$ is the transmit power of the $m$-th IoT device, and $\sigma^2$ is the noise power spectral density at the UAV.
{Note that the transmit power of the $m$-th IoT device is constrained by its maximum transmit power $p_m^{\max}$, i.e., $p_m [n] \leq p_m^{\max}$.}
Hence, the latency and energy consumption for local model uploading can be expressed as
\begin{equation}
    T_m^{\mathrm{up}} [n] = \frac{d_m [n]}{R_m^{\rm{up}} [n]},
\end{equation}
\begin{equation}
    E_{m}^{\mathrm{up}} [n] = p_m[n]T_m^{\mathrm{up}}[n].
\end{equation}

In this work, we adopt the synchronous federated averaging scheme \cite{mcmahan2017communication}.
Thus, the total latency for the $m$-th IoT device to wait the completion of local model uploading of other IoT devices in time slot $n$ can be expressed as
\begin{equation} 
\begin{split}
&T_m^{\mathrm{wait}} [n] \\
&=  \max_{k \in \mathcal{M}} \left \{ T_k^{\mathrm{train}} [n] + T_k^{\mathrm{up}} [n] \right \} 
- \left ( T_m^{\mathrm{train}} [n] + T_m^{\mathrm{up}} [n]\right ).
\end{split}
\end{equation}

\subsection{UAV Model}
After receiving the local models from all IoT devices, the UAV starts model aggregation. 
Let $f_{\mathrm{UAV}} [n]$ denote the computing frequency of the UAV, and $\phi$ denote the computation resources required to process one bit of data, {where the computing frequency of the UAV is constrained by the maximum computing frequency $F_{\max}$, i.e., $f_{\mathrm{UAV}} [n] \leq F_{\max}$.}
Then, the latency and energy consumption for model aggregation can be given by 
\begin{equation}
T_{\mathrm{UAV}}^{\mathrm{agg}} [n] =\frac{\phi \sum_{m=1}^M d_m[n] }{f_{\mathrm{UAV}}[n]},
\end{equation}
\begin{equation}
    E_{\mathrm{UAV}}^{\mathrm{agg}} [n] = \kappa (f_{\mathrm{UAV}} [n])^3 T_{\mathrm{UAV}}^{\mathrm{agg}} [n],
\end{equation}
where $\kappa$ denotes the effective capacitance of the UAV. 
The aggregated global model is then downloaded to each IoT device over the allocated RBs.
The global model downloading latency for the $m$-th IoT device is expressed as
\begin{equation}
T_m^{\mathrm{down}} [n] = \frac{d_m^{\mathrm{agg}}[n]}{R_m^{\mathrm{down}} [n]},
\end{equation}
where
$d_m^{\mathrm{agg}}[n]$ denotes the size of the global model updates, and
$R_m^{\mathrm{down}} [n]$ is the model downloading rate from the UAV to the $m$-th IoT device, which is given by
\begin{equation}
R_m^{\mathrm{down}} [n] = \sum_{u = 1}^{U} \alpha_{m, u} [n] B \log_2 \left(1 + \frac{p_{{\mathrm{u}},m}[n] g_m [n]}{  B \sigma_m^2}\right), 
\end{equation}
where $p_{{\mathrm{u}},m}[n]$ denotes the transmit power from the UAV to the $m$-th IoT device, {which is constrained by the maximum power $p_{{\mathrm{u}}}^{\max}$, i.e., $p_{{\mathrm{u}},m}[n] \leq p_{{\mathrm{u}}}^{\max}$,} 
and $\sigma_m^2$ denotes the noise power spectral density at the $m$-th IoT device.
Moreover, the energy consumption for model downloading to the $m$-th IoT device can be expressed as
\begin{equation}
E_m^{\mathrm{down}} [n] =  p_{{\mathrm{u}},m}[n] T_m^{\mathrm{down}} [n].
\end{equation}
The overall latency of the $n$-th training round is thus given by
\begin{equation}
\begin{split}
    T_{\mathrm{FL}} [n] =\ & 
    \max_{k \in \mathcal{M}} \left \{ T_k^{\mathrm{train}} [n] + T_k^{\mathrm{up}} [n] \right \} \\ &+ T_{\mathrm{UAV}}^{\mathrm{agg}} [n] +
    \max_{k \in \mathcal{M}} \left \{ T_k^{\mathrm{down}} [n] \right \}.
\end{split}
\end{equation}
We assume that the UAV has the maximum velocity $V_{\max}$ and maximum acceleration $a_{\max}$ \cite{10606316}, and the constraints can be expressed as
\begin{equation}
    \| \mathbf{v} [n] \| = \frac{ \| \mathbf{q}[n] - \mathbf{q} [n-1] \| } {T_{\mathrm{FL}} [n-1]} \leq V_{\max} ,\ n \backslash \{1\},
\end{equation}
\begin{equation}
    \| \mathbf{a} [n] \| = \frac{ \| \mathbf{v}[n] - \mathbf{v} [n-1] \| } {T_{\mathrm{FL}} [n-1]} \leq a_{\max} ,\ n \backslash \{1,2\}.
\end{equation}

Meanwhile, given the kinematic requirements, the UAV must maintain a minimum speed $V_{\min}$ to ensure the flight, which is written as
\begin{equation}
     V_{\min} \leq \| \mathbf{v} [n] \|,\ n \backslash \{1\}.
\end{equation}

\subsection{Problem Formulation}
\begin{figure}[!htbp]
	\centering
	\includegraphics[width=1\linewidth]{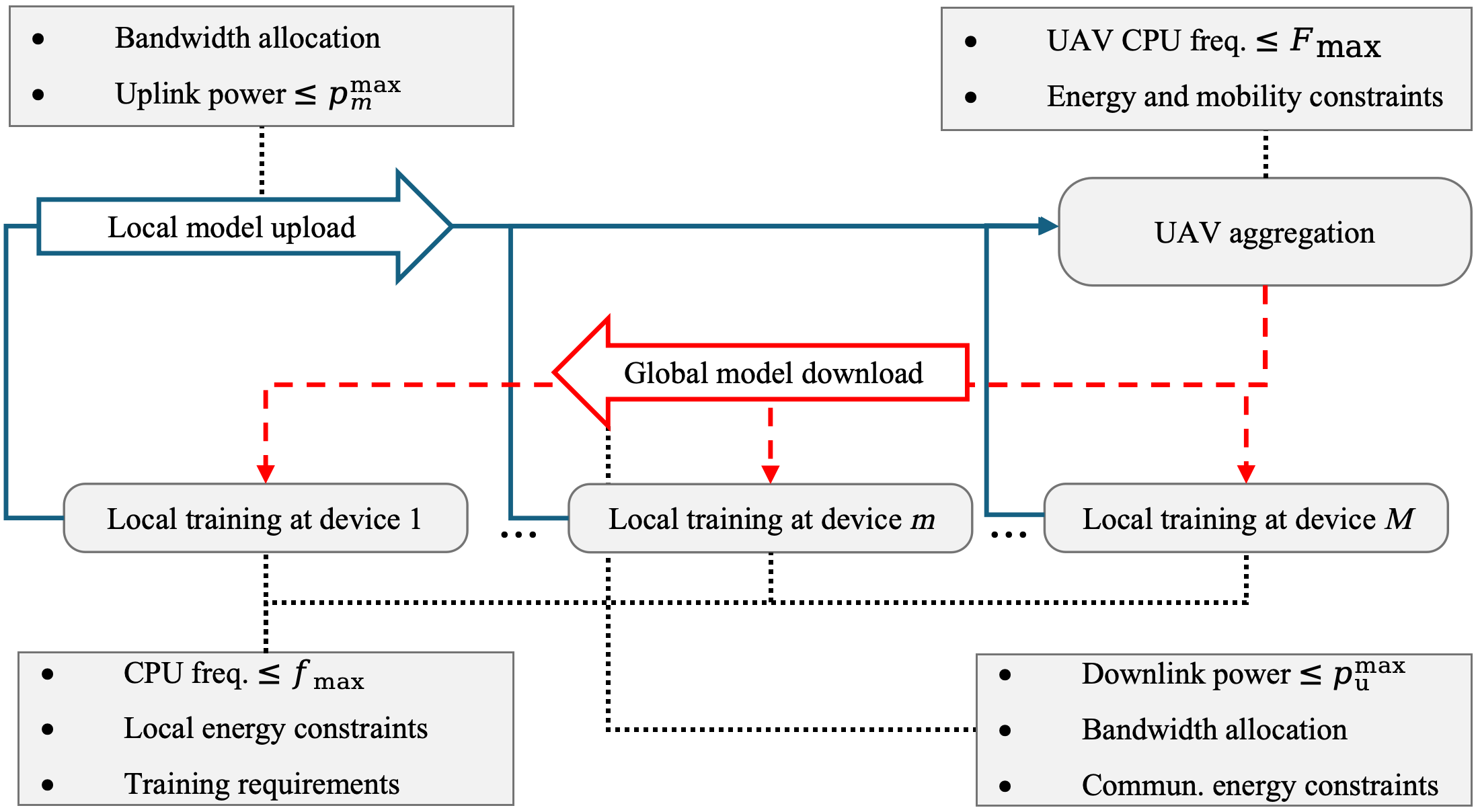}
	\caption{{The FL training process and the constraints among the UAV and the IoT devices.}}
	\label{fig:work}
\end{figure}

In this work, we aim to minimize the total latency of all IoT devices over $N$ time slots. To achieve this, we jointly optimize the allocation of RBs $\boldsymbol{\alpha} \triangleq \{ \alpha_{m,u}[n], \forall m, u, n \}$, the computing frequencies of both IoT devices and the UAV $\mathbf{f} \triangleq \{ f_m[n], f_{\mathrm{UAV}} [n], \forall m, n \}$, the transmit powers of both IoT devices and the UAV $\mathbf{p} \triangleq \{ p_m[n], p_{{\mathrm{u}},m} [n], \forall m, n \}$, and the trajectory of the UAV $\mathbf{Q} \triangleq \{ \mathbf{q}[n], \forall n \}$. The optimization problem can be formulated as
\begin{subequations}\label{P}
\begin{flalign}
\min_{\boldsymbol{\alpha} , \mathbf{f}, \mathbf{p}, \mathbf{Q} } \ & 
 \sum_{m=1}^{M}  \sum_{n=1}^{N} L_m[n]  \label{PA}\\
 {\rm{s.t.}}  \quad & \mathbf{q}[1] = \mathbf{q}_{\mathrm{i}}, \quad \mathbf{q}[N] = \mathbf{q}_{\mathrm{f}},   \label{Pb}\\
& \| \mathbf{v} [n] \| \leq V_{\max} ,
\ \forall n \backslash \{1\},\label{Pc}\\
& \| \mathbf{a} [n] \| \leq a_{\max} ,
\ \forall n \backslash \{1,2\},\label{Pd}\\
& V_{\min} \leq \| \mathbf{v} [n] \|,
\ \forall n \backslash \{1\}, \label{Pe}\\
& \alpha_{m,u} [n] \in \left \{ 0,1 \right \},  
\forall m, u, n, \label{Pg}\\
& \sum_{m \in \mathcal{M}} \alpha_{m,u} [n] \leq 1,\ 
\forall u, n, \label{Ph}\\
& 1 \leq \sum_{u \in \mathcal{U}} \alpha_{m,u} [n],\ 
\forall m, n, \label{Pi}\\
&{0 \leq f_m [n] \leq f_{\max},\ \forall m, n, \label{Pj-11}}\\
&{0 \leq f_{\mathrm{UAV}} [n] \leq F_{\max},\ \forall n, \label{Pj-12}}\\
&{0 \leq p_m [n] \leq p_m^{\max},\ \forall m, n, \label{Pj-1}}\\
&{0 \leq p_{{\mathrm{u}},m}[n] \leq p_{{\mathrm{u}}}^{\max},\ \forall m, n, \label{Pj-3}}\\
& \sum_{n \in \mathcal{N}} \left(E_m^{\mathrm{train}} [n] + E_m^{\mathrm{up}} [n] \right) \leq E_m^{\max},\ 
\forall m, \label{Pj}\\
& \sum_{n \in \mathcal{N}} \left( E_{\mathrm{UAV}}^{\mathrm{agg}} [n] + \sum_{m \in \mathcal{M}}E_m^{\mathrm{down}} [n] \right) \leq E_{\mathrm{UAV}}^{\max}, \label{Pk}
\end{flalign}
\end{subequations} 
where $L_m[n]$ denotes the latency of the $m$-th IoT device in time slot $n$, which is given by
\begin{equation}
    L_m[n] \triangleq T_m^{\mathrm{train}} [n] + T_m^{\mathrm{up}} [n] + T_m^{\mathrm{wait}} [n] + T_{\mathrm{UAV}}^{\mathrm{agg}} [n] + T_m^{\mathrm{down}} [n].
\end{equation}
$E_m^{\max}$ and $E_{\mathrm{UAV}}^{\max}$ represent the maximum energy budgets for the $m$-th IoT device and the UAV, respectively.
The constraints (\ref{Pb})-(\ref{Pe}) represent the kinematic limitations of the UAV, while constraints (\ref{Pg})-(\ref{Pi}) specify the RB allocation restrictions.
Furthermore,
{constraints \eqref{Pj-11} and \eqref{Pj-12} represent the computing frequency limitations for IoT devices and the UAV, respectively.}
{constraints \eqref{Pj-1} and \eqref{Pj-3} represent the transmit power limitations for IoT devices and the UAV, respectively.}
Constraints (\ref{Pj}) and (\ref{Pk}) establish the energy consumption limits for IoT devices and the UAV, respectively.
{The FL training and the constraints among the UAV and the IoT devices are shown in Fig.~\ref{fig:work}.}

\section{Joint Optimization of Resource Allocation and UAV Trajectory}
In this section, we first discuss the convergence of the FL system.
Based on this, we propose an efficient AO algorithm to solve the problem \eqref{P}.

\subsection{FL Convergence Analysis}
Firstly, we give the following assumptions that are widely adopted in the convergence analysis of FL \cite{9210812, 9264742, 9170917, 9606731, 10032291, 10220154, 10051719, 10500333}.

\begin{assumption}
    ($L$-smooth) The loss function of each IoT device $\mathcal{L}(\boldsymbol{\theta}_m [n]; \mathcal{D}_m [n])$ is $L$-Lipschitz continuous gradient with respect to $\boldsymbol{\theta}_m$, which is given by
    \begin{equation}
        \Vert \nabla \mathcal{L}(\boldsymbol{\theta}_m^r; \mathcal{D}_m) - \nabla \mathcal{L}(\boldsymbol{\theta}_m^{r-1}; \mathcal{D}_m) \Vert \leq
        L \Vert \boldsymbol{\theta}_m^r - \boldsymbol{\theta}_m^{r-1} \Vert,
    \end{equation}
where the time slot index is omitted for simplicity, and $L$ is determined by the loss function.
\label{assumption1}
\end{assumption}

\begin{assumption}
    (Strong convexity) The loss function of each IoT device $\mathcal{L}(\boldsymbol{\theta}_m [n]; \mathcal{D}_m [n])$ is strongly convex, i.e.,
    \begin{equation}
    \begin{split}
        \mathcal{L}(\boldsymbol{\theta}_m^r; \mathcal{D}_m) \geq\ &
        \frac{\Upsilon}{2}
        \Vert \boldsymbol{\theta}_m^r - \boldsymbol{\theta}_m^{r-1} \Vert^2 + \mathcal{L}(\boldsymbol{\theta}_m^{r-1}; \mathcal{D}_m)\\
        &+ \left ( \boldsymbol{\theta}_m^r - \boldsymbol{\theta}_m^{r-1}
        \right )^{\mathsf{T}} \nabla \mathcal{L}(\boldsymbol{\theta}_m^r; \mathcal{D}_m),
    \end{split}
    \end{equation}
    where $\Upsilon$ is a constant determined by the loss function.
    \label{assumption2}
\end{assumption}

Under Assumption \ref{assumption1} and \ref{assumption2}, the lower bound for the local training rounds, and the lower bound for the total communication rounds
for ensuring the convergence of FL are presented in the following Proposition \ref{theorem1} and \ref{theorem1-2}, respectively.

\begin{proposition}
\label{theorem1}
    Given the local accuracy $\epsilon_{\mathrm{Local}}$, the local learning rate $\upsilon \leq \frac{2}{L}$,
    the lower bound of local training rounds $\check{R}$ can be given by
    \begin{equation}
        \check{R} =
        \frac{2}{\upsilon\Upsilon (2-L\upsilon)}\ln \left(\frac{1}{\epsilon_{\mathrm{Local}}}\right).
    \label{localround}
    \end{equation}
    \begin{proof}
        {Please refer to Appendix \ref{app0-local}.}
    \end{proof}
\end{proposition}
\begin{proposition}
\label{theorem1-2}
    Given the global accuracy $\epsilon_{\mathrm{UAV}}$, and set $0 \leq \varrho \leq \frac{\Upsilon}{L}$,
    the lower bound of global UAV aggregation rounds $\check{N}$ can be given by
    \begin{equation}
        \check{N} =
        \frac{2L^2}{\Upsilon^2 \varrho (1-\epsilon_{\mathrm{Local}})} \ln \left( \frac{1}{\epsilon_{\mathrm{UAV}}}\right).
        \label{uavround}
    \end{equation}
    \begin{proof}
        {Please refer to Appendix \ref{app0-uav}.}
    \end{proof}
\end{proposition}

\subsection{Problem Transformation}
Problem (\ref{P}) is a typical non-convex mixed-integer nonlinear programming (MINLP) problem due to the high degree of coupling between the discrete variables $\boldsymbol{\alpha}$ and the continuous variables $\mathbf{f,p}$ and $\mathbf{Q}$. Moreover, the constraints (\ref{Pe}), (\ref{Pj}), and (\ref{Pk}) are non-convex.
 Currently, there is no general algorithm that can effectively solve such problems. 
To address this problem, we first transform the optimization objective by introducing an auxiliary variable  $\boldsymbol{\eta} = \{\eta[n], \forall n\}$, in which $\eta[n] = \max_{m \in \mathcal{M}} \{ T_m^{\mathrm{train}}[n] + T_m^{\mathrm{up}}[n] \}$ represents the maximum training and uploading latency among all IoT devices in time slot $n$. Then, the total latency for the $m$-th IoT device can be rewritten as
\begin{equation}
\mathcal{L}_m [n] = \eta [n] + T_{\mathrm{UAV}}^{\mathrm{agg}} [n] + T_m^{\mathrm{down}} [n].
\end{equation}
Accordingly, the problem (\ref{P}) can be rewritten as 
\begin{subequations}\label{P2}
\begin{flalign}
\min_{\boldsymbol{\alpha} , \mathbf{f}, \mathbf{p}, \mathbf{Q}, \boldsymbol{\eta} } \  &  \sum_{m=1}^{M}  \sum_{n=1}^{N} \mathcal{L}_m [n] \\
 {\rm{s.t.}} \quad & \eqref{Pb}-\eqref{Pk},\nonumber \\
& T_m^{\mathrm{train}} [n] + T_m^{\mathrm{up}} [n] \leq \eta [n],\ \forall m, n.\label{P2c}
\end{flalign}
\end{subequations} 

Next, we decompose problem \eqref{P2} into three subproblems: the RB allocation, the UAV trajectory optimization, and communication and computation resource allocation.


\subsection{RB Allocation}
Given the computation resource and power allocation of IoT devices and the UAV $\{\mathbf{f, p}\}$, and the trajectory of the UAV $\{ \mathbf{Q}\}$, the problem (\ref{P2}) can be simplified as
\begin{subequations}\label{P3}
\begin{flalign}
  \min_{ \boldsymbol{\alpha}, \boldsymbol{\eta}} \ \ \ &  
  \sum_{m=1}^{M}  \sum_{n=1}^{N} \mathcal{L}_m [n]  \\
 {\rm{s.t.}}  \quad & \eqref{Pg}-\eqref{Pi}, \eqref{Pj}, \eqref{Pk}, \eqref{P2c}. \nonumber
\end{flalign}
\end{subequations}
Problem (\ref{P3}) is an MINLP problem with binary variables. To handle this issue, we relax the integer constraints (\ref{Pg}) by
\begin{equation}\label{P3_1}
    0 \le \alpha_{m,u} [n] \le 1, \forall m, u, n.
\end{equation}
Based on the above analysis, problem \eqref{P3} becomes
\begin{subequations}\label{P3A}
\begin{flalign}
 \min_{ \boldsymbol{\alpha}, \boldsymbol{\eta}  } \ \ \ & 
  \sum_{m=1}^{M}  \sum_{n=1}^{N} \mathcal{L}_m [n]  \\
 {\rm{s.t.}}  \quad &\eqref{Ph},\eqref{Pi}, \eqref{Pj}, \eqref{Pk}, \eqref{P2c}, \eqref{P3_1}. \nonumber
\end{flalign}
\end{subequations}
It can be verified that the objective function and all constraints are convex with respect to $\alpha_{m,u} [n]$ and $\eta[n]$.
Therefore, problem (\ref{P3A}) is convex and can be efficiently solved using the CVX tool. To gain deeper insights, we derive the closed-form solution of problem (\ref{P3A}) using the Karush-Kuhn-Tucker (KKT) conditions, as presented in the following Theorem \ref{kkto1}.
\begin{theorem} \label{kkto1}
The optimal solution to problem \eqref{P3A} can be expressed as
\begin{equation}
    \begin{split}
\alpha_{m,u}^{*}[n] = \left. \sqrt{ \frac{ \mathcal{W}_{m,u}[n] }{ \Tilde{\varphi}_{u,n} - \Tilde{\varpi}_{m,n}}} \ \right|_{0} ^{1}, \label{alphaopt}
    \end{split}
\end{equation}

\begin{equation}
\eta[n]^{*} = \max_{m \in \mathcal{M}} \left( T_m^{\mathrm{train}} [n] + \frac{d_m [n]}{R_m [n](\alpha_{m,u}^{*}[n])}  \right).
\end{equation}
where $\left.a\right|_{b} ^{c}=\min \{\max \{a, b\}, c\}$ and
\begin{equation}
\begin{split}
\mathcal{W}_{m,u}[n] = & \ \frac{\Tilde{\mu}_{m,n}d_m [n] + \Tilde{\gamma} p_m[n] d_m [n]}{ B \log_2 \left(1 + \frac{p_m[n] g_m[n]}{ B \sigma^2}\right)} \\
& +\frac{d_m^{\mathrm{agg}}[n] + \Tilde{\xi}_{m} p_{{\mathrm{u}},m}[n] d_m^{\mathrm{agg}}[n]}{ B \log_2 \left(1 + \frac{p_{{\mathrm{u}},m}[n] g_m [n]}{  B \sigma_m^2}\right)}  .
\end{split}
\end{equation}

\begin{proof}
Please refer to Appendix \ref{appendice1}.
\end{proof}
\end{theorem}
According to (\ref{alphaopt}), the value of $\alpha_{m,u}^{*}[n]$ depends on $\mathcal{W}_{m,u}[n]$ for given Lagrange multipliers $ \boldsymbol{\Tilde{\varphi}}$, $\boldsymbol{\Tilde{\varpi}}$,
$\mathbf{\Tilde{\xi}}$, $\mathbf{\Tilde{\gamma}}$, and $\mathbf{\Tilde{\mu}}$ that are associated with the constraints (\ref{Ph})-(\ref{Pk}), and \eqref{P2c}. Here, we provide an explanation for each term in $\mathcal{W}_{m,u}[n]$. The first term is defined as the latency and energy expended by the $m$-th IoT device in selecting the $u$-th RB to upload local model parameters to the UAV. The second term is defined as the latency and energy expended by the UAV in downloading to the $m$-th IoT device over the $u$-th RB.

\begin{remark}
Noting that the optimal solution $\mathbf\alpha^{*}$ to the problem \eqref{P3A} is within the range from $0$ to $1$, rather than the deterministic binary value of $0$ or $1$ to the original problem \eqref{P3}.
Therefore, if the optimal solution $\mathbf\alpha^{*}$ obtained from the problem \eqref{P3A} is a binary value, the relaxation is tight, and the solution $\mathbf\alpha^{*}$ is also a feasible solution to the original problem \eqref{P3}. Otherwise, we need to reconstruct a feasible solution to the original problem \eqref{P3} based on the optimal solution $\mathbf\alpha^{*}$ to the relaxed problem \eqref{P3A}.
\end{remark}

Accordingly, we utilize
the decomposition approach \cite{8247211} to reconstruct the feasible solution, which is tailored to the considered FL system.
During the reconstruction, we divide each RB into $\chi$ sub-RBs. Consequently, the total number of sub-RBs is $\Tilde{U} = U \times \chi$. Thus, the number of sub-RBs occupied by the $m$-th IoT device in time slot $n$ is denoted as $ \alpha_{m}^{\rm{total}}[n]=\left\lfloor \chi \times \sum_{u \in \mathcal{U}} \alpha_{m,u}[n] \right\rceil $, where $\lfloor x\rceil$ denotes the nearest integer of $x$.
{
We note that the reconstruction process may introduce a performance gap.
However, it decreases with the bandwidth of the sub-RBs decreases.
Therefore, the reconstruction process can always achieve a near-optimal solution to \eqref{P3} \cite{8247211}.}

\subsection{UAV Trajectory}
Given the RB allocation $\{\boldsymbol{\alpha}\}$, and the computation resource and power allocations of IoT devices and the UAV $\{\mathbf{f, p}\}$, the problem (\ref{P2}) can be simplified as
\begin{subequations}\label{P4}
\begin{flalign}
  \min_{ \mathbf{Q},\boldsymbol{\eta}  } \ \ \  & \sum_{m=1}^{M}  \sum_{n=1}^{N} \mathcal{L}_m [n] \label{P4a} \\
 {\rm{s.t.}}  \quad & 
 \eqref{Pb}-\eqref{Pe},
 \eqref{Pj},\eqref{Pk},\eqref{P2c}. \nonumber
\end{flalign}
\end{subequations}
It can be seen that the $R_m^{\mathrm{up}} [n]$ and $R_m^{\mathrm{down}} [n]$ are non-convex
functions with respect to $\mathbf{q} [n]$. Therefore, we introduce some auxiliary variables and use a successive convex approximation (SCA) method to approximate it.
Firstly, we introduce an auxiliary variable $\check{\Delta}_m [n] \leq H_0^2 + \| \mathbf{q} [n] - \mathbf{s}_m \|^2$, then the uploading data rate can be expressed as
\begin{equation}
R_m [n] \leq \sum_{u=1}^{U} \alpha_{m, u} [n] B \log_2 \left(1 + \frac{p_m[n] g_0}{ B \sigma^2 \check{\Delta}_m [n]}\right),
    \label{appofrm}
\end{equation}
where the right-hand-side of (\ref{appofrm}) is convex with respect to $\check{\Delta}_m [n]$. Moreover, by employing the SCA method, we can derive the lower-bound for the right-hand-side of (\ref{appofrm}) by the first order Taylor expansion at the given position $\check{\Delta}_m^{(l)} [n]$ in the $l$-th iteration, which can be given by (\ref{rmsca}).

\begin{figure*}[!t]
\normalsize

\vspace*{-\baselineskip} 
\begin{equation}
\label{rmsca}
\begin{aligned}
&\sum_{u=1}^{U} \alpha_{m, u} [n] B \log_2 \left(1 + \frac{p_m[n] g_0}{ B \sigma^2 \check{\Delta}_m [n]} \right) = 
\sum_{u \in \mathcal{U}} \alpha_{m, u} [n]  B
\left ( \log_2 \left( \check{\Delta}_m [n] + \frac{p_m[n] g_0}{ B \sigma^2} \right ) - \log_2 ( \check{\Delta}_m [n])
\right ) \\
&\quad\quad\quad\quad\quad\ \geq \sum_{u=1}^{U} \alpha_{m, u} [n] B
\left ( \log_2 \left( \check{\Delta}_m [n] + \frac{p_m[n] g_0}{ B \sigma^2} \right ) - \log_2 ( \check{\Delta}_m^{(l)} [n]) - \frac{1}{\check{\Delta}_m^{(l)} [n]\ln{2}} 
\left ( \check{\Delta}_m^{} [n] - \check{\Delta}_m^{(l)} [n]
\right )
\right )
\triangleq \hat{R}_m [n].
\end{aligned}
\end{equation}

\hrulefill
\end{figure*}

Hence, we have $R_m[n] \leq \hat{R}_m [n]$. Similarly, the downlink data rate of the $m$-th device satisfies
\begin{equation}
R_m^{\mathrm{down}} [n]  \leq \hat{R}_m^{\mathrm{down}} [n],
\end{equation}
where $\hat{R}_m^{\mathrm{down}} [n]$ is the first order Taylor expansion of $\sum_{u=1}^{U} \alpha_{m, u} [n] B \log_2 (1 + \frac{p_{{\mathrm{u}},m}[n] g_0}{ B \sigma_m^2 \check{\Delta}_m^{} [n]})$ at the given position $\check{\Delta}_m^{(l)} [n]$ in the $l$-th iteration.

Furthermore, $t_m^{\mathrm{up}} [n]$ and $t_m^{\mathrm{down}} [n]$ in \eqref{P4a} is related to the trajectory. To convert it, we introduce two auxiliary variables $\Tilde{R}_m^{\mathrm{}} [n] = R_m^{\mathrm{}} [n]$, and $\Tilde{R}_m^{\mathrm{down}} [n] = R_m^{\mathrm{down}} [n]$, which satisfy $\Tilde{R}_m^{\mathrm{}} [n] \leq \hat{R}_m^{\mathrm{}} [n]$ and $\Tilde{R}_m^{\mathrm{down}} [n] \leq \hat{R}_m^{\mathrm{down}} [n]$. Then, we have
\begin{equation}
    \Tilde{T}_m^{\mathrm{up}} [n] = 
    \frac{d_m^{\mathrm{}}}{\Tilde{R}_m^{\mathrm{}} [n]},
    \label{t2app}
\end{equation}
\begin{equation}
    \Tilde{T}_m^{\mathrm{down}} [n] = 
    \frac{d_m^{\mathrm{agg}}}{\Tilde{R}_m^{\mathrm{down}} [n]}.
    \label{t2app2}
\end{equation}

By substituting (\ref{t2app}) and \eqref{t2app2} into constraints (\ref{Pj}), (\ref{Pk}), and (\ref{P2c}), we have the following inequalities, i.e.,
\begin{equation}
    \sum_{n=1}^{N} \left(
E_m^{\mathrm{train}} [n] + p_m[n] \frac{d_m^{\mathrm{}}}{\Tilde{R}_m^{\mathrm{}} [n]}
\right) \leq E_m^{\max},\ \forall m,
\label{P4-11}
\end{equation}
\begin{equation}
    \sum_{n=1}^{N} \left(
E_{\mathrm{UAV}}^{\mathrm{agg}} [n] + \sum_{m \in \mathcal{M}} \left( p_{{\mathrm{u}},m}[n] \frac{d_m^{\mathrm{agg}}}{\Tilde{R}_m^{\mathrm{down}} [n]} \right)
\right) \leq E_{\mathrm{UAV}}^{\max},
\label{P4-12}
\end{equation}

\begin{equation}
    T_m^{\mathrm{train}} [n] + \frac{d_m^{\mathrm{}}}{\Tilde{R}_m^{\mathrm{}} [n]} \leq \eta [n],\ \forall m, n.
    \label{P4-14}
\end{equation}

Additionally, $T_{\mathrm{FL}}[n]$ is time-varying and is non-linear with respect to $\mathbf{q}[n]$, which significantly complicates the optimization of the UAV trajectory. To mitigate this challenge, we employ the $l$-th iterative approximation of the UAV trajectory to estimate the variables within $T_{\mathrm{FL}}[n]$ pertinent to the $(l+1)$-th iteration. Thus, $T_{\mathrm{FL}}[n]$ is rewritten as
\begin{equation}
\begin{split}
\Tilde{T}^{(l)}_{\mathrm{FL}} [n] =\ & 
    \max_{k \in \mathcal{M}} \left \{ T_k^{\mathrm{train}} [n] + T_k^{\mathrm{up}} [n](\mathbf{q}^{(l)} [n]) \right \} \\ &+ T_{\mathrm{UAV}}^{\mathrm{agg}} [n] +
    \max_{k \in \mathcal{M}} \left \{ T_k^{\mathrm{down}} [n](\mathbf{q}^{(l)} [n]) \right \}.
\end{split}
\end{equation}
Then, for the constraint (\ref{Pe}), we utilize the first-order Taylor expansion in the $l$-th iteration to approximate the squared term of (\ref{Pe}), which is given by
\begin{equation}
\begin{aligned}
  &\left ( V_{\min} \Tilde{T}^{(l)}_{\mathrm{FL}} [n-1]  \right )^
2  \leq - \| \mathbf{q}^{(l)} [n] - \mathbf{q}^{(l)} [n-1] \|^2
\\&\quad\quad
+ 2 \left ( \mathbf{q}^{(l)} [n] - \mathbf{q}^{(l)} [n-1]
   \right )^{\mathrm{T}} 
   \left ( \mathbf{q} [n] - \mathbf{q} [n-1]
   \right ) ,
   \ \forall n.
\end{aligned}
\label{vminapp}
\end{equation}

Besides, the inequality $\check{\Delta}_m [n] \leq H_0^2 + \| \mathbf{q} [n] - \mathbf{s}_m \|^2$ is non-
convex. Therefore, it can be approximated by
\begin{equation}
\begin{aligned}
    \check{\Delta}_m [n] \leq
   &\ 2 \left ( \mathbf{q}^{(l)} [n] - \mathbf{s}_m [n]
   \right )^{\mathrm{T}} 
   \left ( \mathbf{q} [n] - \mathbf{q}^{(l)} [n]
   \right )
     \\ &+ H_0^2 + \| \mathbf{q}^{(l)} [n] - \mathbf{s}_m [n] \|^2
   ,\ \forall m, n.
\end{aligned}
\label{dcheckapp}
\end{equation}

Based on the above analysis, the problem (\ref{P4}) can be transformed into
\begin{subequations}\label{P4-1}
\begin{flalign}
  \min_{ \mathbf{Q}, \boldsymbol{\eta}, \boldsymbol{\Delta}, \mathbf{R}  }\ \ \  & \sum_{m=1}^{M}  \sum_{n=1}^{N} \Tilde{\mathcal{L}}_m [n] \\
 {\rm{s.t.}}  \quad & 
 \eqref{Pb}-\eqref{Pd},
 \eqref{P4-11}-\eqref{dcheckapp}, \nonumber\\
 &\Tilde{R}_m^{\mathrm{}} [n] \leq \hat{R}_m^{\mathrm{}} [n],\\
 &\Tilde{R}_m^{\mathrm{down}} [n] \leq \hat{R}_m^{\mathrm{down}} [n],
\end{flalign}
\end{subequations} where $\Tilde{\mathcal{L}}_m [n] = 
  \eta [n] + T_{\mathrm{UAV}}^{\mathrm{agg}} [n] + \Tilde{T}_m^{\mathrm{down}} [n]$, 
  $\boldsymbol{\Delta} =$ $ \{ \check{\Delta}_m [n], \forall m,n \}$, $\mathbf{R} = \{\Tilde{R}_m [n] ,  \Tilde{r}_m^{\mathrm{down}} [n] , \forall m,n \}$.
Given a reference point $\mathbf{q}^{(l)} [n]$, the problem (\ref{P4-1}) is transformed into a solvable convex optimization problem that can be solved by CVX. 

\subsection{Communication and Computing Resource Allocation}
Given the trajectory of the UAV $\{\mathbf{Q}\}$, and the RB allocations $\{\boldsymbol{\alpha}\}$, the problem (\ref{P2}) can be simplified as
\begin{subequations}\label{P5}
\begin{flalign}
  \min_{ \mathbf{f},\mathbf{p},\boldsymbol{\eta}  } \ \ \ &  \sum_{m=1}^{M}  \sum_{n=1}^{N} \mathcal{L}_m [n] \\
 {\rm{s.t.}}  \quad & 
 \eqref{Pj-11}-\eqref{Pk}, \eqref{P2c}. \nonumber
\end{flalign}
\end{subequations}


The constraint (\ref{Pj}) is non-convex with respect to $p_m[n]$, which is presented in proposition \ref{proposition1}.

\begin{proposition}\label{proposition1}
Constraint (\ref{Pj}) is a non-convex constraint with respect to $p_m[n]$, while the left term $E_m^{\mathrm{up}} [n]$ can be written in the form of $g(x) = \frac{ax}{\log_2(1+bx)}$, which is a non-convex function.
\begin{proof}
Please refer to Appendix \ref{appendice2}.
\end{proof}
\end{proposition}

To transform the non-convex constraint, we introduce an auxiliary variable $\gamma_{m}[n]$, which can be expressed as
\begin{equation}
    \gamma_{m}[n] = \frac{1}{\log_2 \left(1 + \frac{p_m[n] g_m[n]}{ B \sigma^2}\right)}. \label{gamma1}
\end{equation}
Then, combining Eq. (\ref{rate_up}) and Eq. (\ref{gamma1}), the local model uploading rate of the $m$-th IoT device can be rewritten as 
\begin{equation}
    R_m[n] = \sum_{u =1}^{U} \alpha_{m, u} [n] B \frac{1}{\gamma_{m}[n]}.
\end{equation}
Then, the transmit power of the $m$-th IoT device can be expressed as
\begin{equation}
    p_m[n] = \frac{B \sigma^2}{g_m[n]} \left( \mathrm{e}^{\frac{\ln(2)}{\gamma_{m}[n]}} - 1 \right).
\end{equation}
{By replacing $p_m[n]$ in the constraint (\ref{Pj-1}) with $\gamma_{m}[n]$, we have the following inequality
\begin{equation}
    0 \leq \frac{B \sigma^2}{g_m[n]} \left( \mathrm{e}^{\frac{\ln(2)}{\gamma_{m}[n]}} - 1 \right) \leq p_m^{\max},\ \forall m, n,
    \label{pj-1new}
\end{equation}}
Accordingly, by replacing $R_m[n]$ and $p_m[n]$ in the constraint (\ref{Pj}) with $\gamma_{m}[n]$, we have the following inequality
\begin{equation}
\begin{split}
&\sum_{n=1}^{N} \left(E_m^{\mathrm{train}} [n] + 
\frac{d_m [n] \sigma^2}{g_m[n]\sum_{u =1}^{U} \alpha_{m,u}[n]} \gamma_{m}[n] (\mathrm{e}^{\frac{\ln(2)}{\gamma_{m}[n]}} - 1)
\right) \\
&\leq E_m^{\max},\ 
\forall m.   
\end{split}
\label{energym_old}
\end{equation}
To better comply with the disciplined convex programming rules in CVX, we introduce a new variable $\Xi_{m}[n]$, which subjects to the following constraint
\begin{equation}
    \Xi_{m}[n] \ge \gamma_{m}[n] (\mathrm{e}^{\frac{\ln(2)}{\gamma_{m}[n]}} - 1).\label{xi_up1}
\end{equation}
Accordingly, by subsisting \eqref{xi_up1} into \eqref{energym_old}, we have
\begin{equation}
\sum_{n =1}^{N} \left(E_m^{\mathrm{train}} [n] + 
\frac{d_m [n] \sigma^2}{g_m[n]\sum_{u=1}^{U} \alpha_{m,u}[n]} \Xi_{m}[n] \right) \leq E_m^{\max}.\label{energym}
\end{equation}
Obviously, the constraint (\ref{energym}) is convex.
Meanwhile, the constraint (\ref{xi_up1}) can be expressed in the following convex constraint, which is given by
\begin{equation}
\gamma_{m}[n] \mathrm{e}^{\frac{\ln(2)}{\gamma_{m}[n]}} \le \Xi_{m}[n] + \gamma_{m}[n] ,\label{xi_up2}
\end{equation} 
Similarly, the constraint (\ref{Pk}) can also be transformed into the following two convex constraints
\begin{equation}
\begin{split}
    &\sum_{n \in \mathcal{N}} \left( E_{\mathrm{UAV}}^{\mathrm{agg}} [n] + \sum_{m \in \mathcal{M}}\left( \frac{d_m^{\mathrm{agg}}[n] \sigma_{m}^2}{g_m[n]\sum_{u=1}^{U} \alpha_{m,u}[n]}\Xi_{m}^{\rm{down}}[n]  \right) \right) \\
    &\leq E_{\mathrm{UAV}}^{\max},
\end{split}
\label{energyuav}
\end{equation}
and
\begin{equation}
\gamma_{m}^{\rm{down}}[n] \mathrm{e}^{\frac{\ln(2)}{\gamma_{m}^{\rm{down}}[n]}} \le \Xi_{m}^{\rm{down}}[n] + \gamma_{m}^{\rm{down}}[n] ,\label{xi_down1}
\end{equation}
where 
\begin{equation}
    \gamma_{m}^{\rm{down}}[n] = \frac{1}{\log_2 \left(1 + \frac{p_{{\mathrm{u}},m}[n] g_m [n]}{  B \sigma_m^2}\right)},\label{gamma2}
\end{equation}
and $\Xi_{m}^{\rm{down}}[n] $ is an auxiliary variable.

{
Similarly, the transmit power of the UAV to communicate with the $m$-th IoT device can be rewritten as
\begin{equation}
    p_{\mathrm{u},m}[n] = \frac{B\sigma_m^2}{g_m[n]} \left( \mathrm{e}^{\frac{\mathrm{ln}(2)}{\gamma_m^{\rm{down}}[n]}} -1
    \right).
\end{equation}
Accordingly, by replacing $p_{\mathrm{u},m}[n]$ in the constraint \eqref{Pj-3} with $\gamma_{m}^{\rm{down}}[n]$, we have the following reformulated inequality
\begin{equation}
    0 \leq \frac{B\sigma_m^2}{g_m[n]} \left( \mathrm{e}^{\frac{\mathrm{ln}(2)}{\gamma_m^{\rm{down}}[n]}} -1
    \right) \leq p_{{\mathrm{u}}}^{\max},\ \forall m, n,
    \label{pj-2new}
\end{equation}
}

Besides, by replacing $R_m[n]$ and $p_m[n]$ in the constraints  (\ref{P2c}) with $\gamma_{m}[n]$, we have the following inequalities
\begin{equation}
T_m^{\mathrm{train}} [n] + \frac{d_m [n] \gamma_{m}[n]}{\sum_{u=1}^{U} \alpha_{m, u} [n] B } \leq \eta [n],\ \forall m, n. \label{Teta}
\end{equation}

Based on the analysis above, the problem (\ref{P5}) can be transformed into
\begin{subequations}\label{P5-1}
\begin{flalign}
  \min_{ \mathbf{f},\boldsymbol{\eta},\boldsymbol{\gamma},\mathbf{\Xi}  } \ \ \ &  \sum_{m=1}^{M}  \sum_{n=1}^{N} \hat{\mathcal{L}}^{}_m [n] \\
 {\rm{s.t.}}  \quad & 
 \eqref{Pj-11},\eqref{Pj-12},\eqref{pj-1new},\eqref{energym}-\eqref{xi_down1}, \eqref{pj-2new},\eqref{Teta}. \nonumber
\end{flalign}
\end{subequations} 
where $\hat{\mathcal{L}}^{}_m [n] =  \eta [n] + T_{\mathrm{UAV}}^{\mathrm{agg}} [n] + \frac{d_m^{\mathrm{agg}}[n] \gamma_{m}^{\rm{down}}[n]}{\sum_{u=1}^{U} \alpha_{m, u} [n] B }$, 
$\boldsymbol{\gamma} = \{ \gamma_{m}[n], $ $\gamma_{m}^{\rm{down}}[n], \forall m,n \}$, and $\mathbf{\Xi} = \{ \Xi_{m}[n],\Xi_{m}^{\rm{down}}[n], \forall m,n \}$.
It can be verified that the objective function and all constraints are convex.
Therefore, problem (\ref{P5-1}) is convex and can be efficiently solved using the CVX tool.


\subsection{Algorithm Analysis}
The proposed algorithm for solving problem (\ref{P})
is provided in Algorithm \ref{Algorithm1}, which alternately optimizes the RB allocation, the UAV trajectory, and communication and computation
resource allocation in an iterative manner until the objective value converges or the maximum iteration number is reached.
The following two propositions analyze the convergence and complexity of Algorithm \ref{Algorithm1}.

\begin{algorithm} \footnotesize
\caption{The proposed algorithm for solving problem (\ref{P})}
\label{Algorithm1}
\begin{algorithmic}
\REQUIRE {An initial feasible point $\{\boldsymbol{\alpha}^{0}, \mathbf{Q}^{0}, \mathbf{f}^{0}, \mathbf{p}^{0} \}$;}\\
\textbf{Initialize:} Iteration number $l=0$, precision threshold $\varepsilon$, and number of maximum iterations $l_{\max}$;\\
\REPEAT
\STATE Solve the problem (\ref{P3A}) to get the RB allocation strategy $\boldsymbol{\alpha}^{l+1}$ for given $\mathbf{Q}^{l}, \mathbf{f}^{l}, \mathbf{p}^{l} $;\\
\STATE Solve the problem (\ref{P4-1}) to get the UAV trajectory $\mathbf{Q}^{l+1}$ for given $\boldsymbol{\alpha}^{l+1}, \mathbf{f}^{l}, \mathbf{p}^{l} $;\\
\STATE Solve the problem (\ref{P5-1}) to get  communication and computation
resource allocation strategy $\mathbf{f}^{l+1}, \mathbf{p}^{l+1}$ for given $\mathbf{Q}^{l+1}, \boldsymbol{\alpha}^{l+1} $;\\
\STATE Update the objective function value according to above variables, i.e. $\Phi\left(\boldsymbol{\alpha}^{l+1}, \mathbf{Q}^{l+1}, \mathbf{f}^{l+1}, \mathbf{p}^{l+1}\right)$;
\STATE Update $l=l+1$; \\
\UNTIL The objective function between two adjacent iterations is smaller than precision threshold $\varepsilon$ or $l > l_{\max}$;\\
\ENSURE {$\Phi^{*}\left(\boldsymbol{\alpha}^{*}, \mathbf{Q}^{*}, \mathbf{f}^{*}, \mathbf{p}^{*}\right)$, $\boldsymbol{\alpha}^{*}$, $\mathbf{Q}^{*}$, $\mathbf{f}^{*}$, and $\mathbf{p}^{*}$}.\\
\end{algorithmic}
\end{algorithm}


\begin{proposition}
\label{proposition2}
    The objective function of problem \eqref{P} keeps decreasing, as the number of iteration increases. Therefore, Algorithm \ref{Algorithm1} is guaranteed to converge.
\begin{proof}
Please refer to Appendix \ref{appendice3}.
\end{proof}
\end{proposition}

\begin{proposition}
\label{proposition3}
    The overall computational complexity $\Omega$ of Algorithm \ref{Algorithm1} is given by \cite{yang2023delay,meng2022multi}
\begin{equation}
\begin{split}
\Omega =&\ \mathcal{O} \left( \mathcal{L}\left( (MUN+N)^{3.5} + (3MN+3N)^{3.5} ) \right.\right. \\
&\ \left.\left. \quad + (7MN+2N)^{3.5} \right) \log(1/\epsilon_{a}) \right),
\end{split}
\end{equation}
where $\mathcal{L}$ denotes the iteration number of Algorithm \ref{Algorithm1}, and $\epsilon_{a}$ represents the stopping tolerance.
\begin{proof}
Please refer to Appendix \ref{appendice4}.
\end{proof}
\end{proposition}

\section{Numerical Results}
In this section, we evaluate the performance of our proposed scheme.
We consider an FL system with one UAV and $M=20$ IoT devices, which locate in a region of $1,000 \times 1,000$$\mathrm{m}^2$. The UAV is flying at a height of $H = 100\mathrm{m}$ \footnote{
{
The key simulation parameters
are selected based on the following considerations \cite{8956055, 9687317, 9954169, 9951396,10050796, 10180104,10185964} : 1) Regulatory compliance (100m altitude aligns with FAA commercial UAV limits); 2) LoS channel models (urban 100m air-to-ground links ensure strong LoS); 3) Computational tractability (200 slots balance trajectory design, FL training, and complexity); 4) IoT benchmarks (20 devices match smart city sensor densities). These key parameters reflect industry-standard UAV deployments while enabling reproducibility across academic and industrial implementations.}
}.
Other simulation parameters are summarized in
Table \ref{t1}.
For performance comparison, we consider four benchmark schemes:
1) {frequency division multiple access (FDMA) scheme,} which equally allocates the bandwidth to all IoT devices;
2) {fixed trajectory scheme,} in which the trajectory of the UAV is fixed;
3) {fixed user allocation scheme,} where the computing frequency and transmit power allocation of all IoT devices are fixed;
4) {thresholding scheme,}
which utilizes a pre-defined threshold $\delta$ to reconstruct the RB allocation.
Specifically, if $\alpha_{m,u}^{*}[n]$ is greater than $\delta$, we set $\alpha_{m,u}^{*}[n]$ to $1$; otherwise, we set $\alpha_{m,u}^{*}[n]$ to $0$.

\begin{table}[!htbp] \scriptsize
\centering
\caption{Main parameter settings.}
\label{t1}
\begin{tabular}{p{5.650cm}c}
\toprule
\centering
Parameters& Value\\
\midrule  %
\centering The number of time slots, $N$& $200$\\
\centering The communication bandwidth, $B$& $1 \ \rm{MHz}$\\
\centering The number of resource blocks, $U$& $30$\\
\centering The density of the Poisson distribution, $\rho_{\mathrm{sensed}}$ and $\rho_{\mathrm{dropped}}$& $50$\\
\centering The data size of local model updates, $d_m[n]$& $[3.5-4.5]*10^{5} \rm{bits}$\\
\centering CPU cycles per bit calculation at each IoT device, $\phi_m$& $100 \ \rm{cycles/bit}$\\
\centering CPU cycles per bit calculation at the UAV, $\phi$& $100 \ \rm{cycles/bit}$\\
\centering The noise spectral density, $\sigma^2,\sigma_{m}^2$& $-174$ $\rm{dBm/Hz}$\\
\centering The energy efficiency coefficient of local computation, $\kappa$& $10^{-28} $\\
\centering The energy efficiency coefficient of model aggregation, $\kappa_{\rm{UAV}}$& $10^{-28} $\\
\centering The energy capacity of IoT devices, $E_m^{\max}$& $\leq 100$ $\rm{J}$\\
\centering The energy capacity of the UAV, $E_{\rm{UAV}}^{\max}$& $\leq 2000$ $\rm{J}$\\
\bottomrule
\end{tabular}
\end{table}

\subsection{Convergence of Algorithm \ref{Algorithm1}}
\begin{figure}[!htbp]
	\centering
\includegraphics[width=0.84\linewidth]{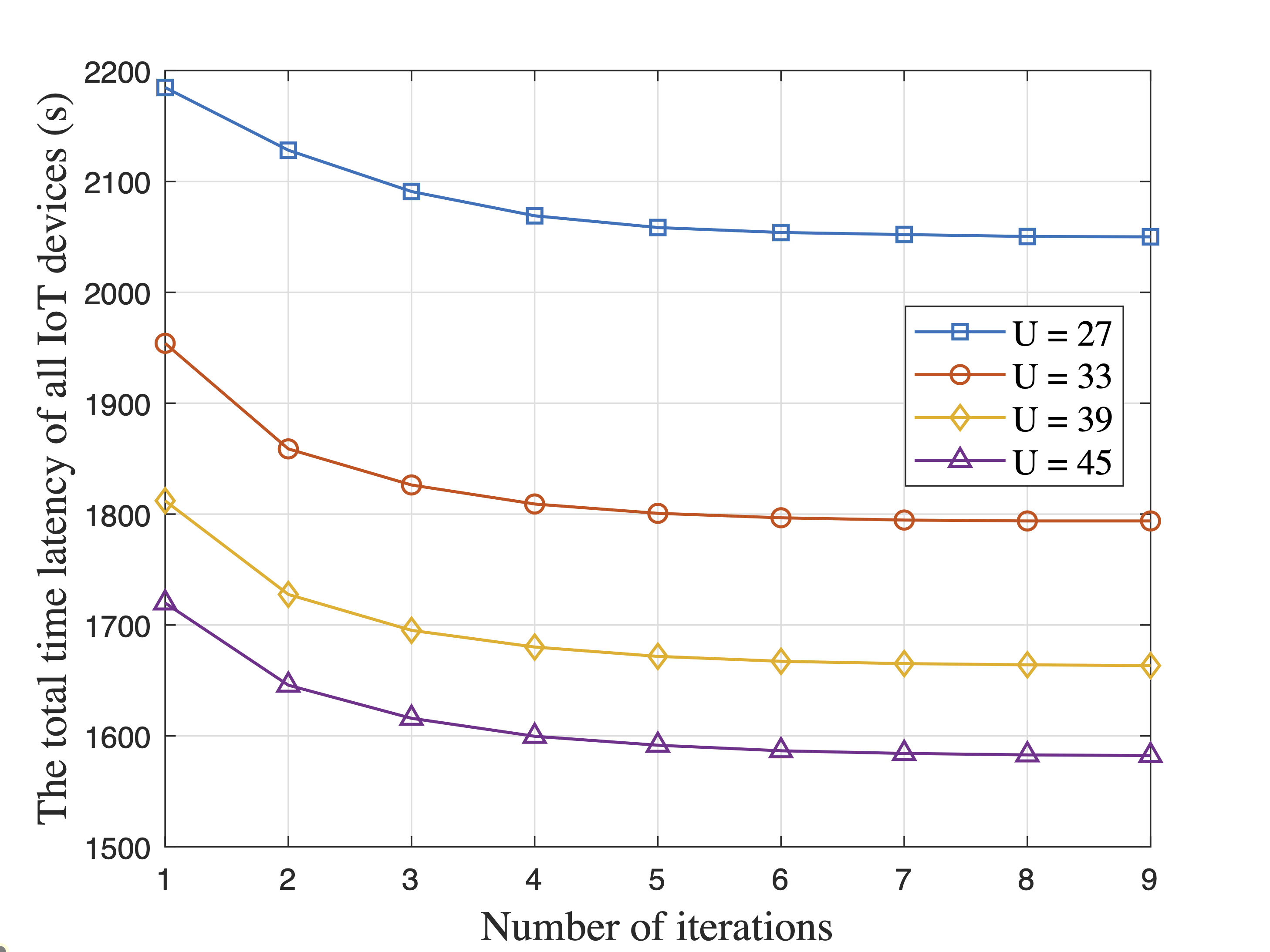}
	\caption{The total latency under different number of RBs versus the number of iterations.}
	\label{fig:convergence}
\end{figure}
Fig. \ref{fig:convergence} presents the total latency of all IoT devices under different number of RBs with respect to the number of iterations.
The curves show that the total latency consistently decreases at the beginning, and gradually converges, thus confirming the convergence analysis in Proposition \ref{proposition2}.
Notably, as the number of RBs increases, the total latency continues to decrease, which can be attributed to the enhanced model uploading and downloading rates with reduced transmission latency.

\subsection{UAV Trajectory}
\begin{figure}[!htbp]
	\centering
	\includegraphics[width=0.84\linewidth]{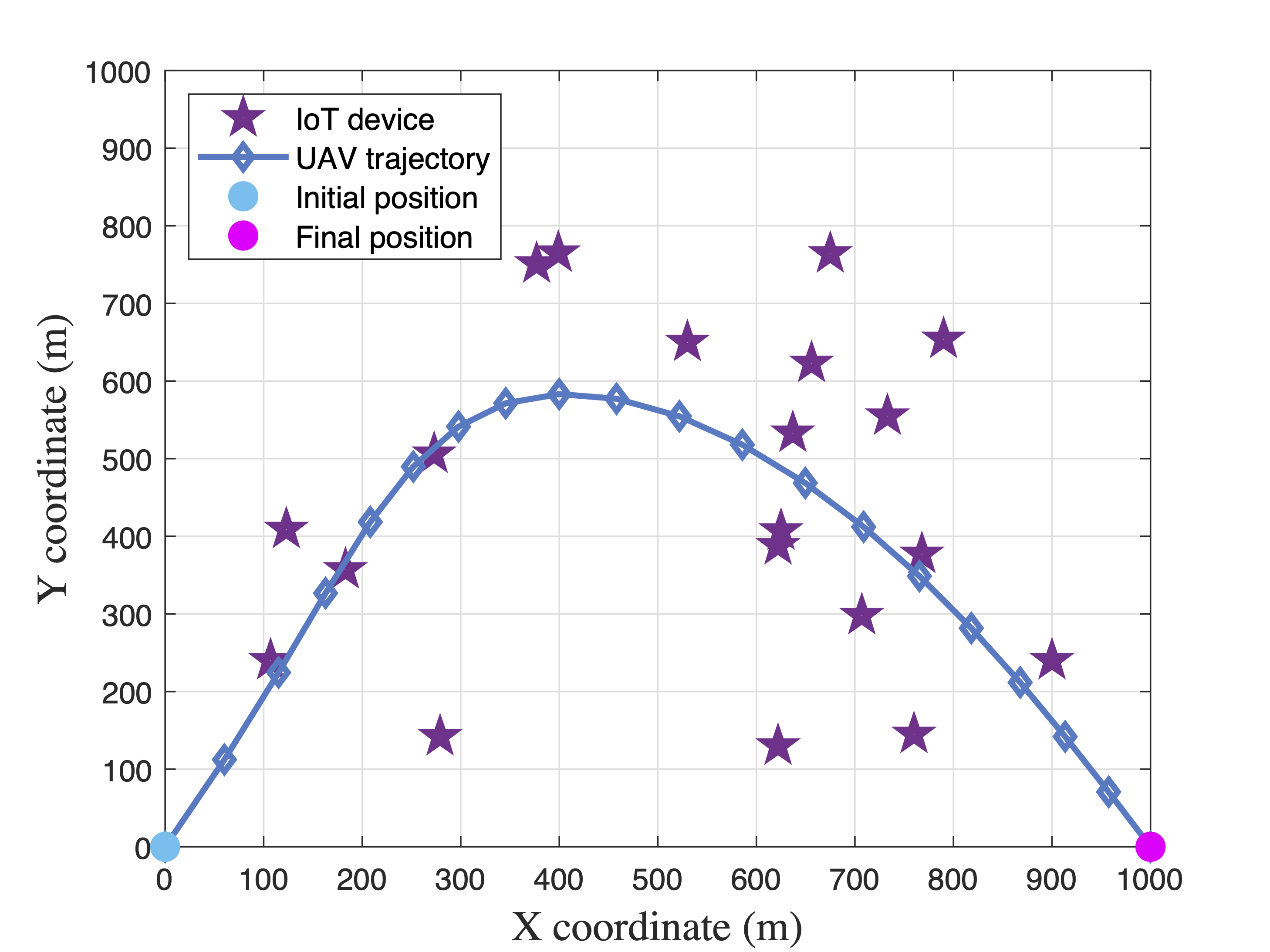}
	\caption{The optimized trajectory of the UAV.}
	\label{fig:trajectory2}
\end{figure}
Fig. \ref{fig:trajectory2} illustrates the optimized trajectory of the UAV from a 2D perspective.
The trajectory is sampled every $10$ rounds and the sampled points are marked with blue hollow $\Diamond$.
The location of each IoT devices is marked with purple solid $\star$.
It can be observed that the UAV first passes through the dense area of IoT devices on the eastern side of the region. Upon reaching the center, it changes direction and flies southwest towards the mission destination.
This trajectory is optimized to minimize the total latency of all IoT devices, taking into account the energy consumption,
computing frequency, and transmit power allocation of the UAV and IoT devices as well as the current position of the UAV.

\subsection{Overall Latency}
\begin{figure}[!htbp]
	\centering
	\includegraphics[width=0.84\linewidth]{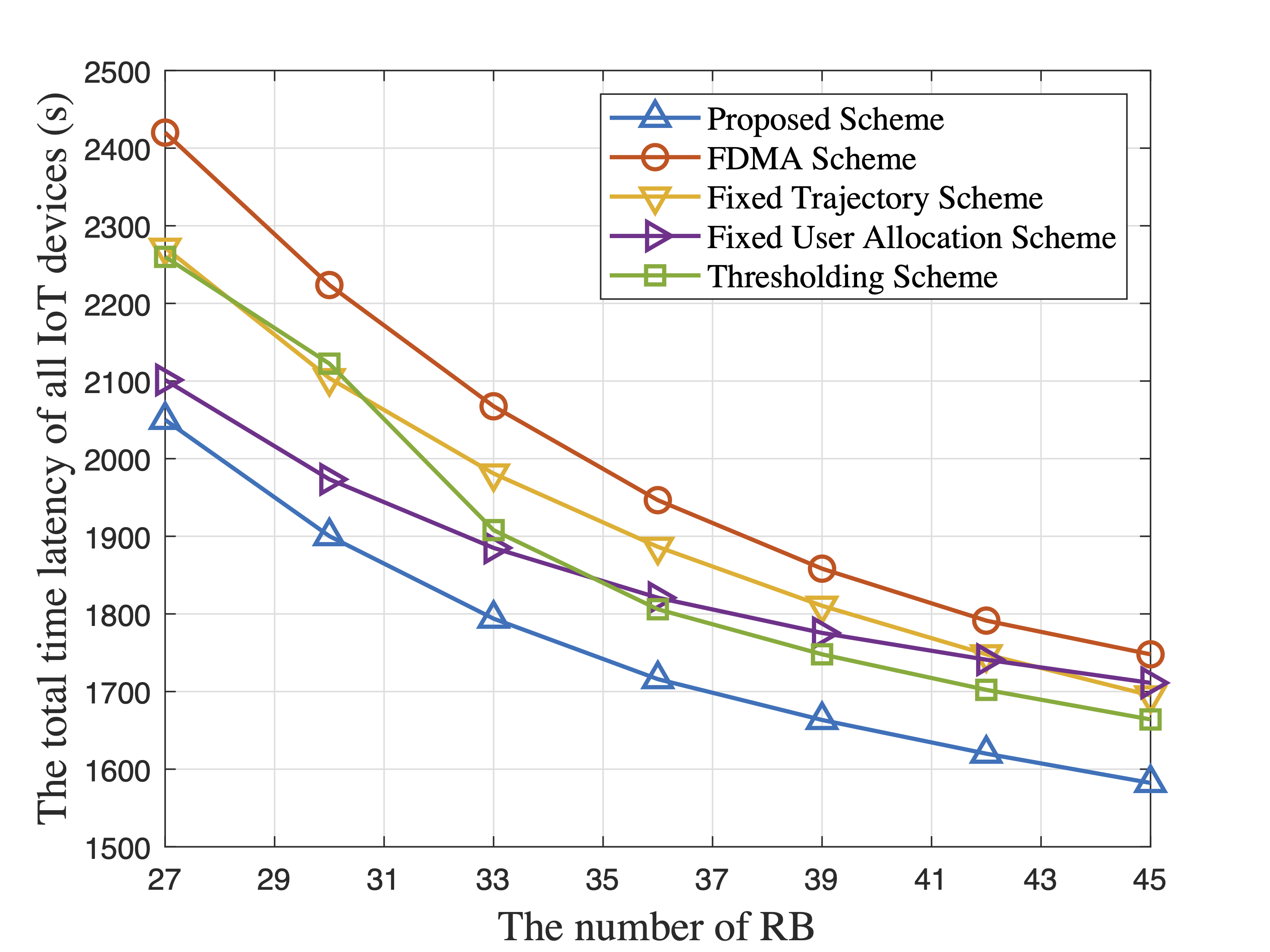}
	\caption{The total latency versus 
 the number of RBs.}
	\label{fig:numberRB}
\end{figure}
Fig. \ref{fig:numberRB} shows the total latency of all IoT devices versus the number of RBs.
We can observe that compared to other benchmark schemes, the proposed scheme achieves the lowest latency. The reason is that the proposed scheme can simultaneously optimize the allocation of RBs, transmit power, computing frequency, and the UAV trajectory, based on the communication quality between the UAV and each IoT device. As the number of RBs increases, the latency for all schemes decreases, which can be attributed to the increased bandwidth resources for improving the data rates among the UAV and IoT devices.

\begin{figure}[!htbp]
	\centering
\includegraphics[width=0.84\linewidth]{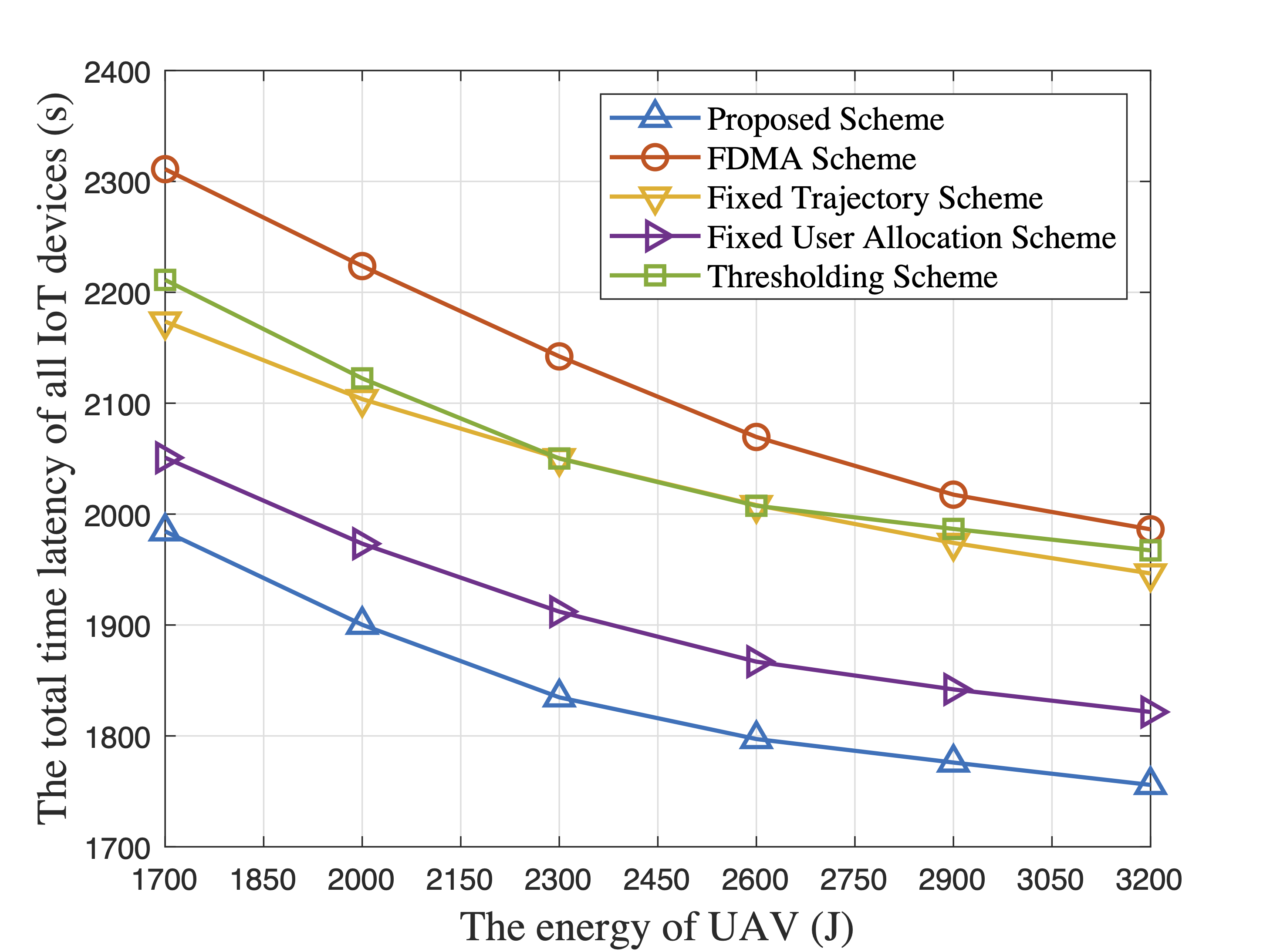}
	\caption{The total latency versus the energy reserve of the UAV.}
	\label{fig:energy}
\end{figure}
The total latency under different FL energy reserves of the UAV is presented in Fig. \ref{fig:energy}.
As the energy reserves of the UAV increase, the latency of all schemes decrease, while our proposed scheme always achieves the minimum latency. The reason is that the increased energy reserves allow the UAV to use higher transmit power during model downloading, while meeting the total energy consumption requirement. This leads to reduced training latency for all IoT devices.
Since the proposed scheme effectively optimizes the maneuver of the UAV and the resource allocation, it consistently maintains the lowest latency as the energy reserve increases.

\begin{figure}[!htbp]
	\centering
	\includegraphics[width=0.84\linewidth]{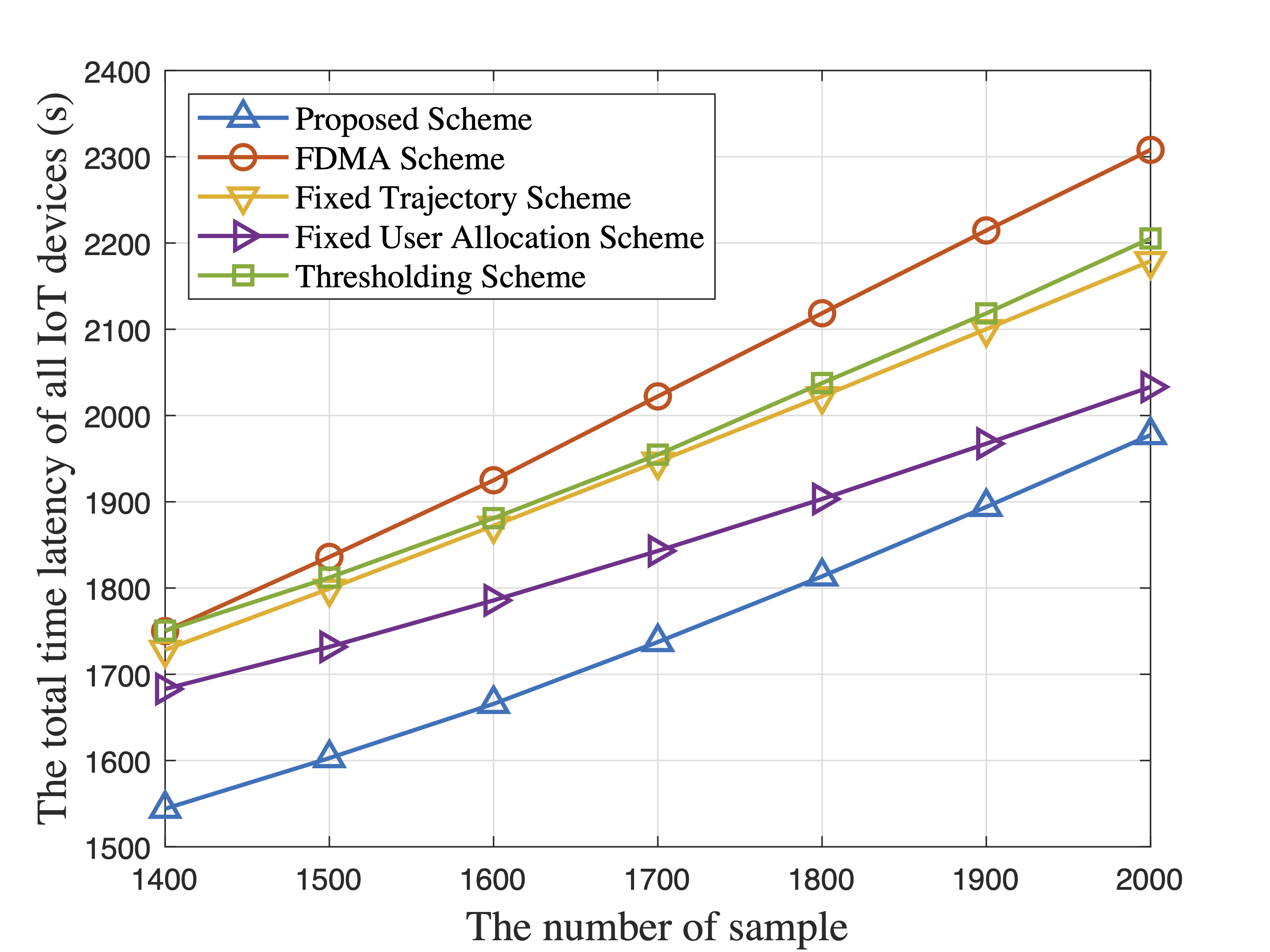}
	\caption{The total latency versus the number of samples.}
	\label{fig:numbersample}
\end{figure}
\begin{figure}[!htbp]
	\centering
	\includegraphics[width=0.84\linewidth]{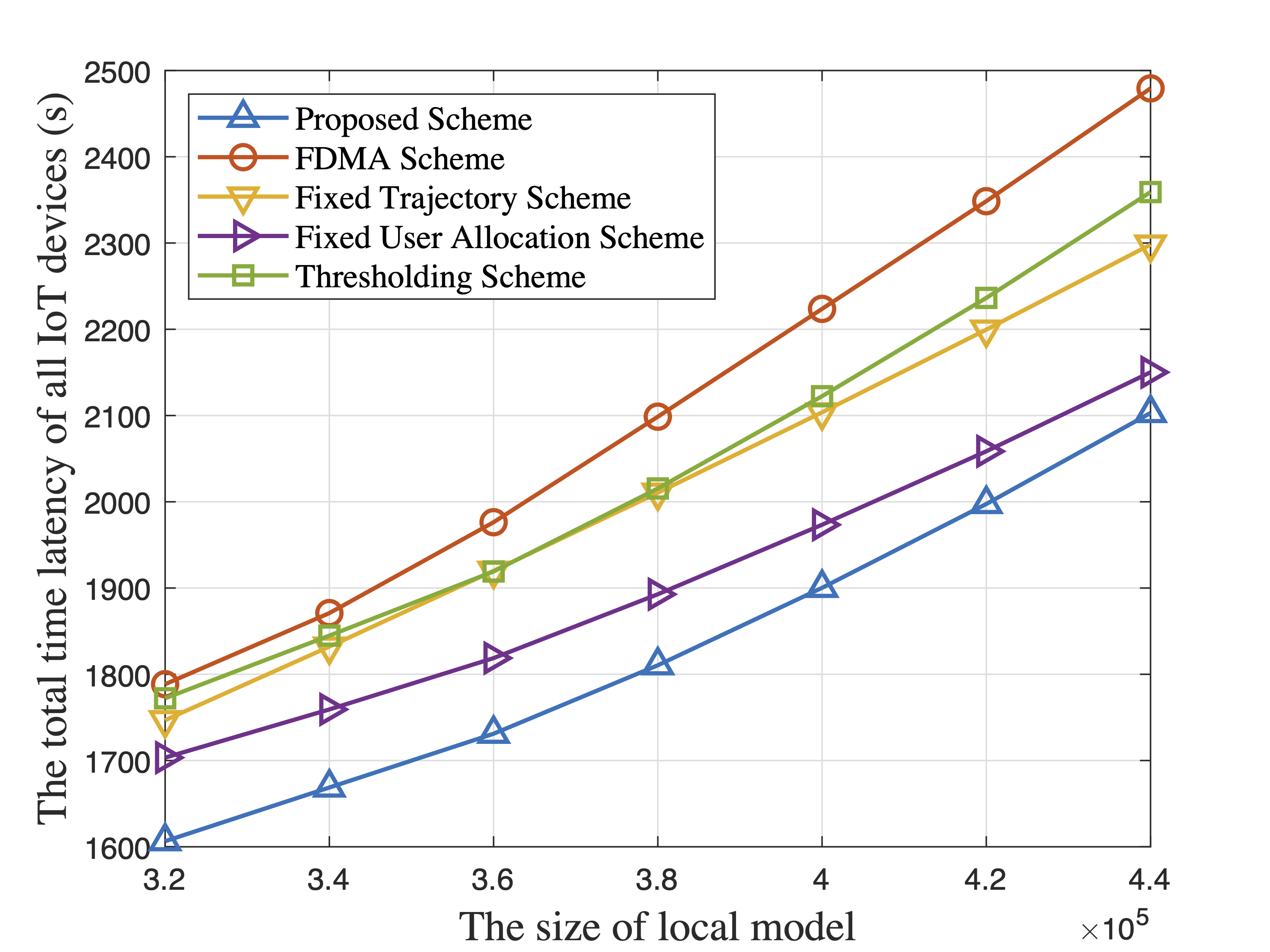}
	\caption{The total latency versus the local model size.}
	\label{fig:local model}
\end{figure}

Fig. \ref{fig:numbersample} and Fig. \ref{fig:local model} present the total latency versus the number of samples, and the local model size of each IoT device, respectively.
As the number of local samples increases, the latency of all schemes increases. The reason is that more local samples leads to more computations for local training, while IoT devices are constrained by their limited computing capability.
Besides, our proposed scheme still outperforms all benchmark schemes in terms of latency.
Furthermore, as illustrated in Fig. \ref{fig:local model}, the latency of all schemes increase as the local model size increases.
The reason is that a larger local model results in increased local model uploading latency.
Similarly, our scheme always achieves the lowest latency.

\subsection{Convergence of FL}

\begin{figure}[!htbp]
	\centering
	\includegraphics[width=0.84\linewidth]{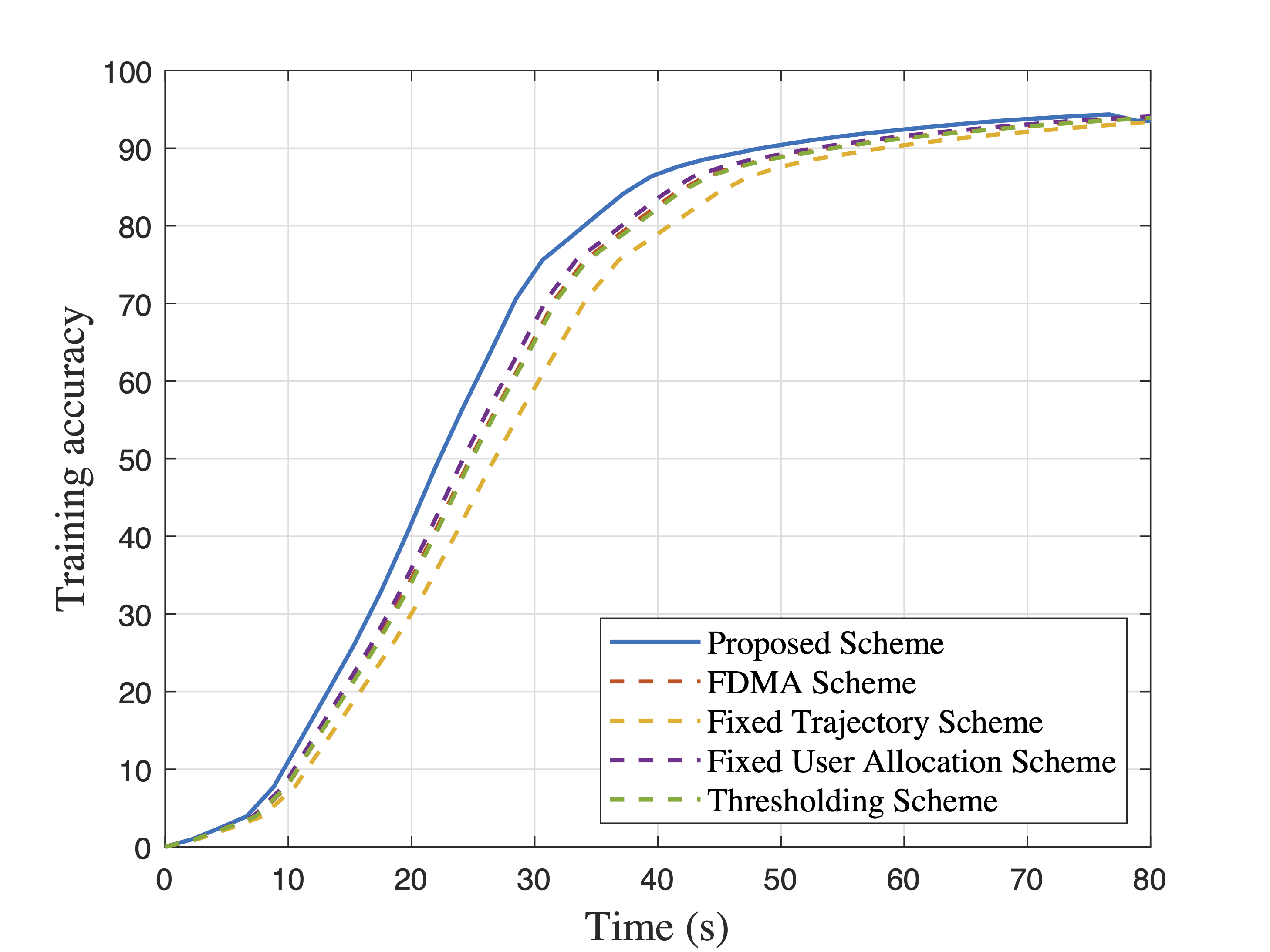}
	\caption{Training accuracy over total training latency of different schemes.}
	\label{fig:acc}
\end{figure}

\begin{figure}[!htbp]
	\centering
	\includegraphics[width=0.84\linewidth]{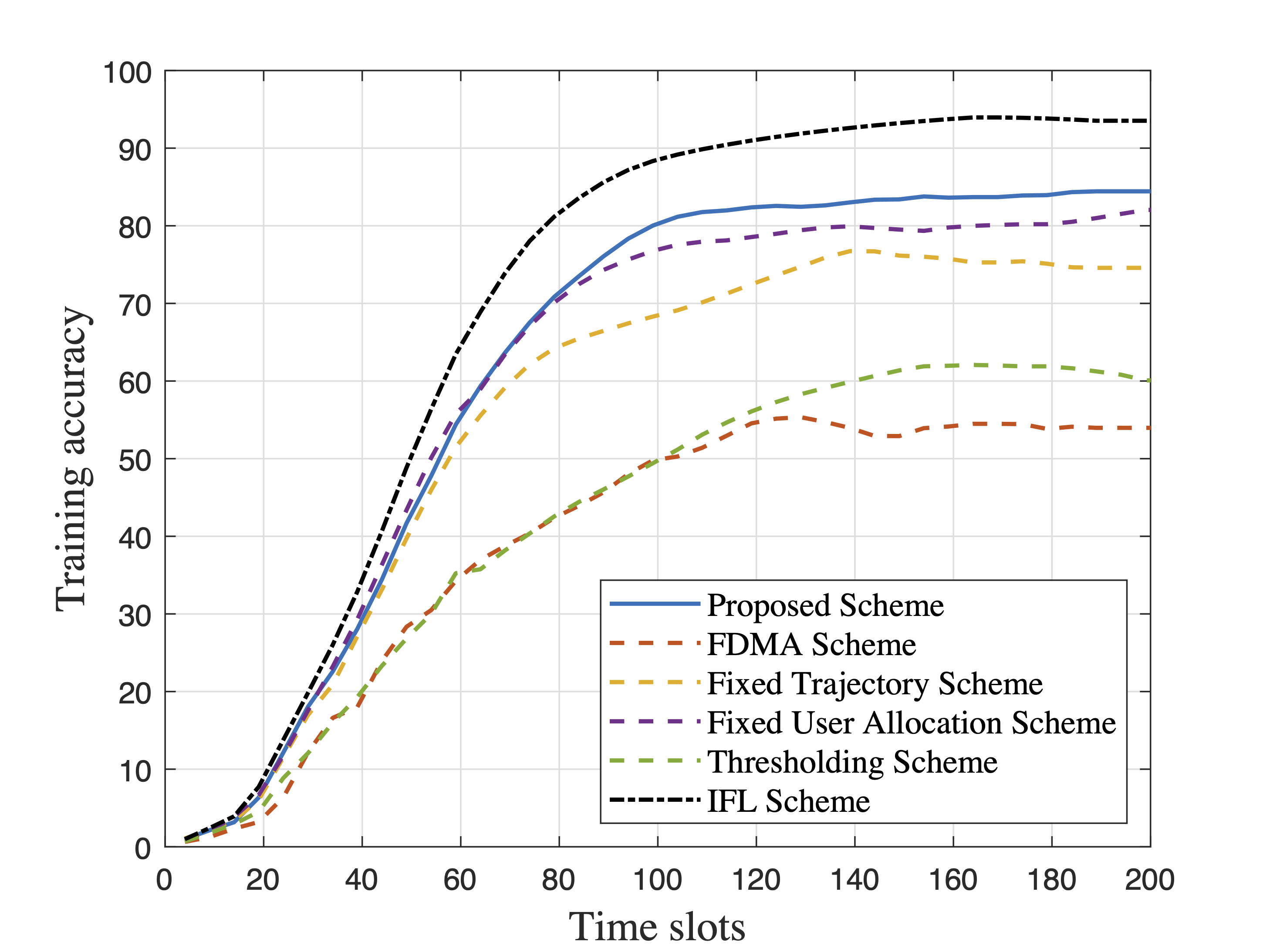}
	\caption{Training accuracy over total time slots of different schemes.}
	\label{fig:acc2}
\end{figure}

We train a neural network on the MINST dataset for a classification task in the UAV-enabled FL system \cite{10518048}.
Each IoT device 
has only two categories of data. Therefore, the training data are non-identical and independent distributed across the devices. The
local dataset for each IoT device is updated across different time slots.
Moreover, we consider two scenarios.
The first scenario involves FL training without restricting the time duration of each time slot, and the result is illustrated in Fig. \ref{fig:acc}.
The second scenario introduces a time slot threshold to ensure latency efficiency, and the result is illustrated in Fig. \ref{fig:acc2}.
In this case, if the total training latency of the $m$-th IoT device exceeds the threshold, the UAV will not aggregate its local model to reduce the total energy consumption.
Besides, we define an ideal scheme with no time slot constraint, namely, ideal FL (IFL) scheme.

Fig. \ref{fig:acc} shows the FL training accuracy over the total training latency of different schemes.
All schemes have sufficient time during each time slot to achieve convergence. We can observe that the convergence speed of the proposed scheme is faster than other benchmark schemes. The reason is that, the proposed scheme can jointly optimize the allocation of communication and computation resources, achieving better latency performance. In contrast, although the other benchmark schemes are able to converge, they consume more communication, computation, and energy resources.

Fig. \ref{fig:acc2} illustrates the FL training accuracy over total time slots of different schemes.
As we can observe, the accuracy of all schemes steadily improves at the beginning of the training process.
After approximately $120$ time slots, the proposed scheme achieves the highest accuracy among all schemes. Moreover, it converges to an accuracy that is closest to that of the IFL scheme.
The reason is that,
the proposed scheme achieves the lower total latency compared to the benchmark schemes. Therefore, the proposed scheme meets the threshold requirement in most time slots, resulting in faster and more effective training performance.

\section{Conclusion}
In this paper, we proposed a UAV-enabled FL system, where the UAV is dispatched to train the FL model together with multiple IoT devices.
Aiming to minimize the total latency of all IoT devices, we formulated an optimization problem to joint optimize the allocation of RBs, the computing frequency and the transmit power of the UAV and IoT devices, and the trajectory of the UAV.
After analyzing the convergence of the FL algorithm, we decomposed the problem into three subproblems and developed an AO-based algorithm to solve it. Thereafter, the convergence and computational complexity of the proposed AO-based algorithm were thoroughly analyzed.
Numerical results confirmed the convergence of the proposed scheme and demonstrated that the proposed scheme not only outperformed other benchmark schemes in terms of latency and training speed, but also achieved the training
speeds that closely approximate the IFL scheme.

\begin{appendices}
    \section{{Proof of the Proposition \ref{theorem1}}}
    \label{app0-local}
    \begin{proof}
    {
    During local training, the local optimization problem of the $m$-th IoT device satisfies,
    \begin{equation}
        \Tilde{\mathcal{L}}(\boldsymbol{\theta}_m^0; \mathcal{D}_m) \geq
        \Tilde{\mathcal{L}}(\boldsymbol{\theta}_m^1; \mathcal{D}_m) \geq
        \cdots \geq
        \Tilde{\mathcal{L}}(\boldsymbol{\theta}_m^r; \mathcal{D}_m),
        \label{app0-00}
    \end{equation}
    where the time slot index $n$ are omitted for simplicity. To guide convergence, the iteration of local training with accuracy $\epsilon_{\mathrm{Local}}$ should satisfy
    \begin{equation}
    \begin{split}
        \Tilde{\mathcal{L}}(\boldsymbol{\theta}_m^r; \mathcal{D}_m) &- \Tilde{\mathcal{L}}(\boldsymbol{\theta}_m^*; \mathcal{D}_m) \\ &\leq \epsilon_{\mathrm{Local}}\left(\Tilde{\mathcal{L}}(\boldsymbol{\theta}_m^0; \mathcal{D}_m) - \Tilde{\mathcal{L}}(\boldsymbol{\theta}_m^*; \mathcal{D}_m)\right),
    \end{split}
    \label{app0-0}
    \end{equation}
    where $\boldsymbol{\theta}_m^*$ denotes the optimal parameter of the $m$-th IoT device. From \eqref{localupdate}, it follows that \cite{9264742}
    \begin{equation}
    \begin{split}
        \Tilde{\mathcal{L}}(\boldsymbol{\theta}_m^{r}; \mathcal{D}_m) \leq
        \Tilde{\mathcal{L}}(\boldsymbol{\theta}_m^{r-1}; \mathcal{D}_m) - \upsilon \Bar{\mathcal{L}}_m^{r-1}
        + \frac{L \upsilon^2}{2} \Bar{\mathcal{L}}_m^{r-1},
    \end{split}
    \label{app0-1}
    \end{equation}
    where $\Bar{\mathcal{L}}_m^{r-1} = \Vert \nabla \Tilde{\mathcal{L}}(\boldsymbol{\theta}_m^{r-1}; \mathcal{D}_m) \Vert^2$.
    According to \textit{Lemma 5} in \cite{9264742}, we have
    \begin{equation}
        \Upsilon \left(
        \Tilde{\mathcal{L}}(\boldsymbol{\theta}_m^r; \mathcal{D}_m) - \Tilde{\mathcal{L}}(\boldsymbol{\theta}_m^*; \mathcal{D}_m)
        \right) \leq
        \Bar{\mathcal{L}}_m^r.
        \label{app0-2}
    \end{equation}
    Under Assumptions \ref{assumption1} and \ref{assumption2}, by combining \eqref{app0-00}, \eqref{app0-0}, \eqref{app0-1}, and \eqref{app0-2}, we obtain
    \begin{equation}
        \begin{split}
            \Tilde{\mathcal{L}}(&\boldsymbol{\theta}_m^{r}; \mathcal{D}_m) - \Tilde{\mathcal{L}}(\boldsymbol{\theta}_m^*; \mathcal{D}_m)
         \leq \Tilde{\mathcal{L}}(\boldsymbol{\theta}_m^{r-1}; \mathcal{D}_m) - \Tilde{\mathcal{L}}(\boldsymbol{\theta}_m^*; \mathcal{D}_m)
         \\&\quad - \left( \upsilon + \frac{L\upsilon^2}{2} \right) \Upsilon
         \left(
            \Tilde{\mathcal{L}}(\boldsymbol{\theta}_m^{r-1}; \mathcal{D}_m) - \Tilde{\mathcal{L}}(\boldsymbol{\theta}_m^*; \mathcal{D}_m)
         \right) 
         \\ &\leq \left( 1-\frac{\upsilon\Upsilon (2-L\upsilon)}{2} \right)
         \left(
            \Tilde{\mathcal{L}}(\boldsymbol{\theta}_m^{r-1}; \mathcal{D}_m) - \Tilde{\mathcal{L}}(\boldsymbol{\theta}_m^*; \mathcal{D}_m)
         \right) 
         \\ &\leq \left( 1-\frac{\upsilon\Upsilon (2-L\upsilon)}{2} \right)^{r}
         \left(
            \Tilde{\mathcal{L}}(\boldsymbol{\theta}_m^{0}; \mathcal{D}_m) - \Tilde{\mathcal{L}}(\boldsymbol{\theta}_m^*; \mathcal{D}_m)
         \right)
         \\ & \leq \mathrm{e}^{\left(-\frac{\upsilon\Upsilon (2-L\upsilon)}{2}\right)r}\left(
            \Tilde{\mathcal{L}}(\boldsymbol{\theta}_m^{0}; \mathcal{D}_m) - \Tilde{\mathcal{L}}(\boldsymbol{\theta}_m^*; \mathcal{D}_m)
         \right).
        \end{split}
    \end{equation}
    To ensure the inequality \eqref{app0-0}, the lower bound of the local training rounds $\check{R}$ should satisfy the following condition, i.e.,
        \begin{equation}
            \begin{split}
                \mathrm{e}^{\left(-\frac{\upsilon\Upsilon (2-L\upsilon)}{2}\right)\check{R}} = \epsilon_{\mathrm{Local}}.
            \end{split}
            \label{app0-3}
        \end{equation}
    By performing a simple transformation to the inequality \eqref{app0-3}, we can obtain \eqref{localround}.
    }
    \end{proof}

    \section{{Proof of the Proposition \ref{theorem1-2}}}
    \label{app0-uav}
    \begin{proof}
    {
    Similarly, the iteration round of the UAV model aggregation should satisfy
    \begin{equation}
    \begin{split}
        {\mathcal{L}}(\boldsymbol{\theta}[n]) - {\mathcal{L}}(\boldsymbol{\theta}^*) \leq \epsilon_{\mathrm{UAV}}\left({\mathcal{L}}(\boldsymbol{\theta}[0]) - {\mathcal{L}}(\boldsymbol{\theta}^*)\right).
    \end{split}
    \end{equation}
    From \eqref{localob}, \eqref{uavaggre}, and Assumptions \ref{assumption1} and \ref{assumption2}, it follows that \cite{9264742}
    \begin{equation}
    \begin{split}
        {\mathcal{L}}(&\boldsymbol{\theta}[n]) \leq
        {\mathcal{L}}(\boldsymbol{\theta}[n-1]) + \frac{1}{M} \sum_{m=1}^M \nabla {\mathcal{L}}(\boldsymbol{\theta}[n-1])^{\mathsf{T}} \Tilde{\theta}_m [n-1] \\
        &\quad\quad\quad\  + \frac{L}{2M^2}\left \Vert \sum_{m=1}^M \Tilde{\theta}_m [n-1] \right \Vert^2 \\
        &\leq  {\mathcal{L}}(\boldsymbol{\theta}[n-1]) + \frac{1}{M\varrho} \sum_{m=1}^M \Bigg ( \Tilde{\mathcal{L}}(\boldsymbol{\theta}_m [n-1]; D_m [n-1]) \\
        &\quad - {\mathcal{L}}(\boldsymbol{\theta}_m [n-1]; D_m [n-1])
        - \frac{\Upsilon}{2} \left \Vert\Tilde{\theta}_m [n-1] \right \Vert^2 \Bigg ) \\
        &\quad +  \frac{L}{2M^2}\left \Vert \sum_{m=1}^M \Tilde{\theta}_m [n-1] \right \Vert^2 \\
        &\leq {\mathcal{L}}(\boldsymbol{\theta}[n-1]) + \frac{1}{M\varrho} \sum_{m=1}^M \Bigg (
        \Tilde{\mathcal{L}}(\boldsymbol{\theta}_m [n-1]; D_m [n-1]) \\
        &\quad - {\mathcal{L}}(\boldsymbol{\theta}_m [n-1]; D_m [n-1])
        - \frac{\Upsilon-L\varrho}{2} \left \Vert\Tilde{\theta}_m [n-1] \right \Vert^2 \Bigg )
        \\ & \leq {\mathcal{L}}(\boldsymbol{\theta}[n-1]) - \frac{1}{M\varrho} \sum_{m=1}^M \Bigg ( \frac{\Upsilon-L\varrho}{2} \left \Vert\Tilde{\theta}_m [n-1] \right \Vert^2 \\
        & \quad + \frac{\Upsilon(1-\epsilon_{\mathrm{local}})}{2} \left \Vert\Tilde{\theta}_m^* [n-1] \right \Vert^2 \Bigg ).
    \end{split}
    \label{app-pro2-ie1}
    \end{equation}
    From Assumptions \ref{assumption1} and \ref{assumption2}, we have
    \begin{equation}
        \frac{1}{L^2} \left \Vert
        \nabla {\mathcal{L}}(\boldsymbol{\theta}_m^* [n]; D_m [n]) - \nabla {\mathcal{L}}(\boldsymbol{\theta} [n]; D_m [n]) \right \Vert^2 \leq \left \Vert\Tilde{\theta}_m^* [n] \right \Vert^2.
        \label{app-pro2-ie2}
    \end{equation}
    Combining \eqref{app-pro2-ie1} and \eqref{app-pro2-ie2}, we have
    \begin{equation}
    \begin{split}
        {\mathcal{L}}(\boldsymbol{\theta}[n]) &\leq
        {\mathcal{L}}(\boldsymbol{\theta}[n-1]) - \frac{1}{M\varrho} \sum_{m=1}^M \Bigg ( \frac{\Upsilon-L\varrho}{2} \left \Vert\Tilde{\theta}_m [n-1] \right \Vert^2 \\
        &\quad + \frac{\Upsilon(1-\epsilon_{\mathrm{local}})}{2L^2} \left \Vert\nabla {\mathcal{L}}(\boldsymbol{\theta}_m^* [n-1]; D_m [n-1]) \right. \\ &\quad \left.
        - \nabla {\mathcal{L}}(\boldsymbol{\theta} [n-1]; D_m [n-1]) \right \Vert^2  \Bigg ).
    \end{split}
    \label{app-pro2-ie3}
    \end{equation}
    When the local training problem \eqref{localob} reach the optimal solution, we have the following condition,
    \begin{equation}
        \nabla {\mathcal{L}}(\boldsymbol{\theta}_m^* [n]; D_m [n]) - \nabla {\mathcal{L}}(\boldsymbol{\theta} [n]; D_m [n]) = - \varrho \nabla {\mathcal{L}}(\boldsymbol{\theta} [n]).
    \label{app-pro2-ie4}
    \end{equation}
    Substituting \eqref{app-pro2-ie4} into \eqref{app-pro2-ie3}, we have
    \begin{equation}
    \begin{split}
        {\mathcal{L}}(\boldsymbol{\theta}[n]) &\leq
        {\mathcal{L}}(\boldsymbol{\theta}[n-1]) - \frac{1}{M\varrho} \sum_{m=1}^M \Bigg ( \frac{\Upsilon-L\varrho}{2} \left \Vert\Tilde{\theta}_m [n-1] \right \Vert^2 \\
        &\quad + \frac{\Upsilon\varrho(1-\epsilon_{\mathrm{local}})}{2L^2} \left \Vert\nabla  {\mathcal{L}}(\boldsymbol{\theta} [n-1]) \right \Vert^2  \Bigg ).\\&
        \leq
        {\mathcal{L}}(\boldsymbol{\theta}[n-1]) - \frac{1}{M\varrho} \sum_{m=1}^M \Bigg ( \frac{\Upsilon-L\varrho}{2} \left \Vert\Tilde{\theta}_m [n-1] \right \Vert^2 \\&
        \quad + \frac{\Upsilon^2\varrho(1-\epsilon_{\mathrm{local}})}{2L^2} \left (  {\mathcal{L}}(\boldsymbol{\theta} [n-1]) - \mathcal{L}(\boldsymbol{\theta}^* ) \right )  \Bigg ).
    \end{split}
    \label{app-pro2-ie5}
    \end{equation}
    Since we set $0 \leq \varrho \leq \frac{\Upsilon}{L}$, i.e., $0 \leq \Upsilon - L\varrho$, by subtracting $\mathcal{L}(\boldsymbol{\theta}^* )$ from both sides of the inequality \eqref{app-pro2-ie5}, \eqref{app-pro2-ie5} can be rewritten as
    \begin{equation}
    \begin{split}
        {\mathcal{L}}&(\boldsymbol{\theta}[n]) - \mathcal{L}(\boldsymbol{\theta}^* ) \leq {\mathcal{L}}(\boldsymbol{\theta}[n-1]) - \mathcal{L}(\boldsymbol{\theta}^* )
        \\& \quad
        -\left( \frac{\Upsilon^2\varrho(1-\epsilon_{\mathrm{local}})}{2L^2}\right) ({\mathcal{L}}(\boldsymbol{\theta}[n-1]) - \mathcal{L}(\boldsymbol{\theta}^* ))\\
        &= \left(1- \frac{\Upsilon^2\varrho(1-\epsilon_{\mathrm{local}})}{2L^2}\right) ({\mathcal{L}}(\boldsymbol{\theta}[n-1]) - \mathcal{L}(\boldsymbol{\theta}^* ))\\
        &\leq \left(1- \frac{\Upsilon^2\varrho(1-\epsilon_{\mathrm{local}})}{2L^2}\right)^n ({\mathcal{L}}(\boldsymbol{\theta}[0]) - \mathcal{L}(\boldsymbol{\theta}^* ))\\
        &\leq \mathrm{e}^{- \frac{\Upsilon^2\varrho(1-\epsilon_{\mathrm{local}})}{2L^2}n} ({\mathcal{L}}(\boldsymbol{\theta}[0]) - \mathcal{L}(\boldsymbol{\theta}^* )).
    \end{split}
    \end{equation}
    Similar to Appendix \ref{app0-local}, the following condition should hold for the lower bound of the global aggregation rounds $\check{N}$, i.e.,
    \begin{equation}
        \mathrm{e}^{- \frac{\Upsilon^2\varrho(1-\epsilon_{\mathrm{local}})}{2L^2}\check{N}} = \epsilon_{\mathrm{UAV}}.
    \end{equation}
    Then, we can derive \eqref{uavround}.
    }
    \end{proof}

\section{Proof of the Theorem \ref{kkto1}}
\label{appendice1}
\begin{proof} 
By dualizing the constraints \eqref{Ph}, \eqref{Pi}, \eqref{Pj}, \eqref{Pk}, and \eqref{P2c}, we obtain the Lagrangian of (\ref{P3A}) as
\begin{equation}
    \begin{split} \label{Lagrangian1}
&\mathcal{J}(\boldsymbol{\alpha}, \boldsymbol{\eta},\boldsymbol{\Tilde{\varphi}}, \boldsymbol{\Tilde{\varpi}}, \boldsymbol{\Tilde{\xi}}, \boldsymbol{\Tilde{\gamma}},
\boldsymbol{\Tilde{\mu}})\\
&= \sum_{m=1}^{M}  \sum_{n=1}^{N}
\mathcal{L}_m [n]
+ \sum_{u=1}^{U}  \sum_{n=1}^{N} \Tilde{\varphi}_{u,n} \left( \sum_{m=1}^{M} \alpha_{m,u} [n] - 1 \right) \\
& + \sum_{m=1}^{M}  \sum_{n=1}^{N} \Tilde{\varpi}_{m,n} \left( 1 -\sum_{u=1}^{U} \alpha_{m,u} [n] \right)\\
& + \sum_{m=1}^{M} \Tilde{\xi}_{m} \left( \sum_{n=1}^{N} \left(E_m^{\mathrm{train}} [n] + E_m^{\mathrm{up}} [n] \right) - E_m^{\max} \right)\\
& + \Tilde{\gamma} \left( \sum_{n=1}^{N} ( E_{\mathrm{UAV}}^{\mathrm{agg}} [n] + \sum_{m \in \mathcal{M}} E_m^{\mathrm{down}} [n] ) - E_{\mathrm{UAV}}^{\max} \right)\\
&+ \sum_{m=1}^{M}  \sum_{n=1}^{N} \Tilde{\mu}_{m,n}(T_m^{\mathrm{train}}[n]+T_m^{\mathrm{up}}[n]-\eta[n]),
    \end{split}
\end{equation}
where $ \boldsymbol{\Tilde{\varphi}} = \{ \Tilde{\varphi}_{u,n}, \forall u,n \} $, $\boldsymbol{\Tilde{\varpi}} = \{\Tilde{\varpi}_{m,n}, \forall m,n \} $,
$\mathbf{\Tilde{\xi}} = \{ \Tilde{\xi}_{m}, \forall m \}$, $\mathbf{\Tilde{\gamma}} = \{ \Tilde{\gamma} \}$, and $\mathbf{\Tilde{\mu}} = \{\Tilde{\mu}_{m,n}, \forall m,n \}$ are the dual multipliers associated with the corresponding constraints (\ref{Ph})-(\ref{Pk}), \eqref{P2c}, respectively. 
Thus, we obtain the dual problem of problem (\ref{P3A}) as
\begin{equation}
\begin{split}
\max_{\boldsymbol{\Tilde{\varphi}}, \boldsymbol{\Tilde{\varpi}}, \boldsymbol{\Tilde{\xi}}, \boldsymbol{\Tilde{\gamma}},
\boldsymbol{\Tilde{\mu}}} = \left\{ \begin{array}{l}
    \min_{\boldsymbol{\alpha}, \boldsymbol{\eta}}  \mathcal{J}(\boldsymbol{\alpha}, \boldsymbol{\eta},\boldsymbol{\Tilde{\varphi}}, \boldsymbol{\Tilde{\varpi}}, \boldsymbol{\Tilde{\xi}}, \boldsymbol{\Tilde{\gamma}},
\boldsymbol{\Tilde{\mu}})   \\
         {\rm{s.t.}} \quad 0 \le \alpha_{m,u} [n] \le 1, \forall m, u, n.
    \end{array}\right. \label{dual}
\end{split}
\end{equation}

Since problem (\ref{P3A}) is convex and satisfies the Slater's condition, there is no duality gap between (\ref{P3A}) and its dual problem (\ref{dual}). 
According to the KKT conditions, by setting the derivatives of $\mathcal{J}(\boldsymbol{\alpha}, \boldsymbol{\eta}, \mathbf{\Tilde{\varphi}}, \mathbf{\Tilde{\varpi}}, \mathbf{\Tilde{\xi}}, \mathbf{\Tilde{\gamma}},  \mathbf{\Tilde{\mu}} , \mathbf{\Tilde{\lambda}} )$ with respect to $\boldsymbol{\alpha}$ to zero, we have
\begin{equation}
\label{eqkkt1}
\begin{split}
\frac{\partial \mathcal{J}}{\partial \alpha_{m,u}[n]} & =  
- \frac{d_m^{\mathrm{agg}}[n]}{\alpha_{m, u}^{2} [n] B \log_2 \left(1 + \frac{p_{{\mathrm{u}},m}[n] g_m [n]}{  B \sigma_m^2}\right)} \\
&\quad - \frac{\Tilde{\xi}_{m} p_{{\mathrm{u}},m}[n] d_m^{\mathrm{agg}}[n]}{\alpha_{m, u}^{2} [n] B \log_2 \left(1 + \frac{p_{{\mathrm{u}},m}[n] g_m [n]}{  B \sigma_m^2}\right)} \\
&\quad - \frac{ \Tilde{\mu}_{m,n}d_m [n]}{\alpha_{m, u}^{2} [n] B \log_2 \left(1 + \frac{p_m[n] g_m[n]}{ B \sigma^2}\right)} \\
&\quad - \frac{\Tilde{\gamma} p_m[n] d_m [n]}{\alpha_{m, u}^{2} [n] B \log_2 \left(1 + \frac{p_m[n] g_m[n]}{ B \sigma^2}\right)} \\
&\quad + \Tilde{\varphi}_{u,n} - \Tilde{\varpi}_{m,n}  \\
& = 0,
\end{split}
\end{equation}
we can obtain the optimal solution to problem (\ref{P3A}) as
\begin{equation}
    \begin{split}
\alpha_{m,u}^{*}[n] = \left. \sqrt{ \frac{ \mathcal{W}_{m,u}[n] }{ \Tilde{\varphi}_{u,n} - \Tilde{\varpi}_{m,n}}} \ \right|_{0} ^{1},
    \end{split}
\end{equation}
where $\left.a\right|_{b} ^{c}=\min \{\max \{a, b\}, c\}$ and
\begin{equation}
    \begin{split}
\mathcal{W}_{m,u}[n] = &\frac{\Tilde{\mu}_{m,n}d_m [n] + \Tilde{\gamma} p_m[n] d_m [n]}{ B \log_2 \left(1 + \frac{p_m[n] g_m[n]}{ B \sigma^2}\right)} \\
&+\frac{d_m^{\mathrm{agg}}[n] + \Tilde{\xi}_{m} p_{{\mathrm{u}},m}[n] d_m^{\mathrm{agg}}[n]}{ B \log_2 \left(1 + \frac{p_{{\mathrm{u}},m}[n] g_m [n]}{  B \sigma_m^2}\right)}  . \\
    \end{split}
\end{equation}
To achieve the optimal $\boldsymbol{\eta}^{*}$, we take the first derivation with respect to $\eta[n]$, considering that (\ref{Lagrangian1}) is a linear function with respect to $\eta[n]$
\begin{equation}
    \begin{split}
\frac{\partial \mathcal{J}}{\partial \eta[n]} = 1 - \Tilde{\mu}_{m,n}.
    \end{split}
\end{equation}
Note that the optimal $\eta[n] = + \infty $ if $1 - \Tilde{\mu}_{m,n} < 0$. To avoid this, we have $ \Tilde{\mu}_{m,n} \leq 1$. Then we can obtain the optimal $\eta[n]^{*}$, which can be expressed as
\begin{equation}
\eta[n]^{*} = \max_{m \in \mathcal{M}} \left( T_m^{\mathrm{train}} [n] + \frac{d_m [n]}{R_m [n](\alpha_{m,u}^{*}[n])}  \right).
\end{equation}
\end{proof}

\section{Proof of the Proposition \ref{proposition1}}\label{appendice2}
\begin{proof}
The second-order derivative of $g(x)$ is expressed as
\begin{equation}
    \frac{d^{2} g(x)}{d x^{2}} = -\ln(2) 
    \frac{(ab^{2}x+2ab)\ln(1+bx)-2ab^{2}x}{(1+bx)^{2}\ln^{3}(1+bx)}.
\end{equation}
Define $\Tilde{g}(x) = (ab^{2}x+2ab)\ln(1+bx)
-2ab^{2}x$.
The second-order derivative of $\Tilde{g}(x)$ is given by
\begin{equation}
    \frac{d^{2} \Tilde{g}(x)}{d x^{2}} = 
    \frac{ab^{4}x}{b^{2}x^{2}+2bx+1}.
\label{appendixb1}
\end{equation}
Clearly, the function $\Tilde{g}(x)$ is convex.
Additionally, from the following Eq. (\ref{appendixb2}), it is evident that the extremum of the function is $x = 0$.
\begin{equation}
    \frac{d \Tilde{g}(x)}{d x} = ab^{2}\ln(1+bx)-\frac{ab^{3}x}{1+bx}.
    \label{appendixb2}
\end{equation}
Therefore, $\Tilde{g}(x) \ge \Tilde{g}(0) = 0$, and $\frac{d^{2} g(x)}{d x^{2}} \le 0$. Hence, the constraint (\ref{Pj}) is non-convex with respect to $p_m[n]$.

\end{proof}

\section{Proof of the Proposition \ref{proposition2}}
\label{appendice3}
\begin{proof}
Let $\boldsymbol{\alpha}^{l}, \mathbf{Q}^{l}, \mathbf{f}^{l}, \mathbf{p}^{l}$ denote the solution in the $l$-th iteration of Algorithm \ref{Algorithm1}, and $\Phi\left(\boldsymbol{\alpha}^{l}, \mathbf{Q}^{l}, \mathbf{f}^{l}, \mathbf{p}^{l}\right)$ denote the objective function.
Given $\mathbf{Q}^{l}, \mathbf{f}^{l}, \mathbf{p}^{l} $, we solve problem \eqref{P3A} to get the RB allocation strategy $\boldsymbol{\alpha}^{l+1}$. Thus, we have
\begin{equation}
     \Phi\left(\boldsymbol{\alpha}^{l}, \mathbf{Q}^{l}, \mathbf{f}^{l}, \mathbf{p}^{l}\right)
\geq \Phi\left(\boldsymbol{\alpha}^{l+1}, \mathbf{Q}^{l}, \mathbf{f}^{l}, \mathbf{p}^{l}\right).
\label{pps1}
\end{equation}
Then, given $\boldsymbol{\alpha}^{l+1}, \mathbf{f}^{l}, \mathbf{p}^{l} $, we solve problem \eqref{P4-1} to get the UAV trajectory $\mathbf{Q}^{l+1}$, and we have
\begin{equation}
     \Phi\left(\boldsymbol{\alpha}^{l+1}, \mathbf{Q}^{l}, \mathbf{f}^{l}, \mathbf{p}^{l}\right)
\geq \Phi\left(\boldsymbol{\alpha}^{l+1}, \mathbf{Q}^{l+1}, \mathbf{f}^{l}, \mathbf{p}^{l}\right).
\label{pps2}
\end{equation}
Accordingly, we solve problem \eqref{P5-1} to get the communication and computation
resource allocation strategy $\mathbf{f}^{l+1}, \mathbf{p}^{l+1}$ for given $\mathbf{Q}^{l+1}, \boldsymbol{\alpha}^{l+1} $. Hence, we have
\begin{equation}
     \Phi\left(\boldsymbol{\alpha}^{l+1}, \mathbf{Q}^{l+1}, \mathbf{f}^{l}, \mathbf{p}^{l}\right)
\geq \Phi\left(\boldsymbol{\alpha}^{l+1}, \mathbf{Q}^{l+1}, \mathbf{f}^{l+1}, \mathbf{p}^{l+1}\right).
\label{pps3}
\end{equation}
Combining \eqref{pps1}, \eqref{pps2} and \eqref{pps3}, we have
\begin{equation}
     \Phi\left(\boldsymbol{\alpha}^{l}, \mathbf{Q}^{l}, \mathbf{f}^{l}, \mathbf{p}^{l}\right)
\geq \Phi\left(\boldsymbol{\alpha}^{l+1}, \mathbf{Q}^{l+1}, \mathbf{f}^{l+1}, \mathbf{p}^{l+1}\right),
\end{equation}
which indicates that the value of the objective function does not increase over iterations. Moreover, the total latency is bounded because of the finite range of the optimization variables. Hence, Algorithm \ref{Algorithm1} is guaranteed to converge.
\end{proof}

\section{Proof of the Proposition \ref{proposition3}}
\label{appendice4}
\begin{proof}
The computational complexity of Algorithm \ref{Algorithm1} is primarily composed of solving problems (\ref{P3A}), (\ref{P4-1}), and (\ref{P5-1}). In problem (\ref{P3A}), the complexity for optimizing the variables $\{\boldsymbol{\alpha}, \boldsymbol{\eta}\}$ is $(MUN+N)^{3.5}$, where $MUN+N$ is the number of variables. 
In problem (\ref{P4-1}), the complexity for optimizing the variables $\{\mathbf{Q}, \boldsymbol{\eta}, \mathbf{d}, \mathbf{R}\}$ is $(3MN+N)^{3.5}$, where $3MN+N$ is the number of variables. 
In problem (\ref{P5-1}), the complexity for optimizing the variables $\{\mathbf{f},\mathbf{p},\boldsymbol{\eta},\boldsymbol{\gamma},\mathbf{\Xi}\}$ is $(7MN+2N)^{3.5}$, where $MUN+N$ is the number of variables.
Therefore, the total computational complexity is $\Omega = \mathcal{O} ( \mathcal{L}( (MUN+N)^{3.5} + (3MN+3N)^{3.5} ) + (7MN+2N)^{3.5} ) \log(1/\epsilon_{a}) )$.
\end{proof}

\end{appendices}

\bibliographystyle{IEEEtran}
\bibliography{myref}

\end{document}